\documentclass[twocolumn,english,superscriptaddress]{revtex4-1}
\usepackage{color}
\usepackage{float}
\usepackage{amsmath}
\usepackage{graphicx}
\usepackage{ulem}
\usepackage{chngcntr}

\makeatletter

\usepackage[caption=false]{subfig}

\begin{document}
\begin{center}
\date{9-3-2020}
\end{center}

\title{Is friction essential for dilatancy and shear jamming in granular matter?}
\author{Varghese Babu}
\thanks{These authors contributed equally.}
\affiliation{Jawaharlal Nehru Centre for Advanced Scientific Research, Jakkur Campus, Bengaluru 560064, India.}
\author{Deng Pan}
\thanks{These authors contributed equally.}
\affiliation{CAS Key Laboratory of Theoretical Physics, Institute of Theoretical Physics, Chinese Academy of Sciences, Beijing 100190, China}
\author{Yuliang Jin}
\email{yuliangjin@mail.itp.ac.cn}
\affiliation{CAS Key Laboratory of Theoretical Physics, Institute of Theoretical Physics, Chinese Academy of Sciences, Beijing 100190, China}
\affiliation{School of Physical Sciences, University of Chinese Academy of Sciences, Beijing 100049, China}
\author{Bulbul Chakraborty}
\affiliation{Martin Fisher School of Physics, Brandeis University, Waltham, MA 02454, USA.}
\author{Srikanth Sastry}
\email{sastry@jncasr.ac.in}
\affiliation{Jawaharlal Nehru Centre for Advanced Scientific Research, Jakkur Campus, Bengaluru 560064, India.}

\captionsetup[subfigure]{labelformat=empty}

\begin{abstract}
{Granular packings display the remarkable phenomenon of {\textsl{dilatancy}} \cite{reynolds1885lvii}, wherein their  volume increases upon shear deformation. Conventional wisdom and previous results suggest that dilatancy, as also the related phenomenon of shear-induced jamming, requires frictional interactions~\cite{peyneau2008frictionless,baity2017emergent}. Here, we investigate the occurrence of dilatancy and shear jamming in frictionless packings. We show that the existence of isotropic jamming densities $\phi_j$
above the minimal density, the J-point density $ \phi_J$~\cite{liu1998nonlinear, o2003jamming},
leads both to the emergence of shear-induced jamming and dilatancy. Packings at $\phi_J$ form a significant threshold state into which systems evolve in the limit of vanishing pressure under constant pressure shear, irrespective of the initial jamming density $\phi_j$. 
While packings for different $\phi_j$
display
equivalent scaling properties under compression~\cite{chaudhuri2010jamming}, they exhibit striking differences in rheological behaviour under shear. The yield stress under constant volume shear increases discontinuously with density  
when $\phi_j > \phi_J$,
contrary to the continuous behaviour in generic packings that jam at $\phi_J$~\cite{liu1998nonlinear,heussinger2009jamming}. }
\end{abstract}

	
\newpage
\maketitle
\newpage
\maketitle

A large variety of familiar materials, made  of macroscopic or mesoscopic constituent particles, may be characterized as {\it granular matter}. Sands, powders and grains are some examples. Given their large sizes, the individual particles (unlike atoms and molecules in a liquid) do not exhibit spontaneous -- {\it Brownian} -- motion, and are thus referred to as being {\it athermal}. They flow in response to externally applied small forces, but at sufficiently high densities or 
applied stresses, cease to flow, or {\it jam} \cite{Cates_1998,liu1998nonlinear}. Density- or stress-driven jamming is of central importance in comprehending a wide variety of complex rheological properties of granular matter, and forms an essential part of a broader understanding of the transition from flowing states of matter to non-flowing or structurally arrested states, including, {\it e. g.},  the glass transition. 

Density-driven jamming, unjamming and yielding of frictionless hard and soft particles have been investigated extensively since the proposal of the jamming phase diagram~\cite{liu1998nonlinear} which has, as originally proposed, a unique density (packing fraction) at $\phi_J$ characterizing the jamming transition at zero temperature and shear stress. 
Since then it has been shown that the jamming density $\phi_j$ is protocol-dependent and therefore not unique~\cite{chaudhuri2010jamming,parisi2010mean,ozawa2012jamming}, satisfying in general  $\phi_j \geq \phi_J$ ~\cite{ozawa2012jamming}.
However, critical behavior associated with jamming, for example the scaling relationship between pressure and density,  $ p \sim (\phi-\phi_j)$,  remains the same, irrespective of $\phi_j$~\cite{chaudhuri2010jamming}. 

An early proposal that shear deformation, besides density, can induce jamming \cite{Cates_1998}, has recently been explored extensively in experimental and theoretical investigations, largely of frictional, but also frictionless hard and soft sphere systems \cite{bi2011jamming,ren2013reynolds,sarkar2013origin,Sarkar2016,kumar2016memory,vinutha2016disentangling,vinutha_kabir,Setoa,Otsuki2017,Otsuki2018,Jineaat6387,Zhao2019}. In shear jamming, the development of an anisotropic contact network under shear leads to the emergence of a state of finite shear stress and pressure, with their ratio 
peaking at a density-dependent characteristic strain \cite{bi2011jamming,ren2013reynolds,sarkar2013origin,Sarkar2016,Setoa,Otsuki2017,Otsuki2018}.

The shear-strain dependent pressure was termed {\it Reynolds pressure} in~\cite{ren2013reynolds}, reflecting the idea that shear jamming occurs because constant volume conditions frustrate the tendency of granular materials to dilate under shear~\cite{reynolds1885lvii}, a phenomenon widely referred to as {\it dilatancy}. With a similar view, impact-driven and shear-driven jamming in dense suspensions have been related to ``frustrated dilatancy'' effects~\cite{Brown_2012,Clavaud_2017}. Shear jamming and dilatancy in frictional granular matter have thus been viewed as two sides of the same coin. We clarify that our discussion and investigation concern behaviour under quasistatic shear, and do not always apply when shear rates are finite.  

Reynolds' dilatancy in granular materials has been extensively investigated,  motivated by the relevance of the phenomena to soil mechanics \cite{kruyt2016micromechanical,rao2008introduction}.
Many available results suggest an intimate relationship between frictional interactions and dilatancy: 
stress-dilatancy relations couple dilatancy and friction between particles \cite{rowe1962stress}. Recent studies indicate that friction is  important for observing shear jamming and dilatancy 
\cite{behringer2018physics}.  Numerical  studies \cite{baity2017emergent,peyneau2008frictionless,azema2015internal} have reported, and experiments  \cite{Clavaud_2017} have also indirectly indicated, the absence of dilatancy in frictionless systems. 

These observations are at variance  with the simple picture suggested by Reynolds~\cite{reynolds1885lvii}, where dilatancy arises  purely from geometric exclusion effects of hard particles, which should therefore be observed also in frictionless systems.  We aim here to resolve this paradox, and demonstrate conditions under which dilatancy emerges naturally in frictionless sphere assemblies. We show that such conditions depend critically on the presence of a line of jamming points at densities $\phi_j$ above $\phi_J$.



In motivating our study, we note that, below $\phi_J$, initially unjammed frictionless sphere assemblies develop structures under shear, with average geometric contact numbers that increase with density, which can be mechanically stabilized by friction~\cite{vinutha2016disentangling}. 
If the unjammed configurations are at densities above $\phi_J$, shear deformations may create contact networks that satisfy the isostatic jamming condition for frictionless packings which are mechanically stable, leading to the possibility of both shear jamming and dilatancy.  Thus, the absence of dilatancy~\cite{peyneau2008frictionless} and shear jamming~\cite{baity2017emergent} in earlier studies could be due to the failure to obtain unjammed initial configurations above $\phi_J$ rather than to the absence of friction.  



We thus investigate assemblies of two models of frictionless spheres prepared to have jamming densities $\phi_j$ above the minimal jamming density at $\phi_J$, explicitly demonstrating the phenomena of shear jamming and dilatancy, and identifying universal features that can be associated to them. 
Both effects diminish as $\phi_{\rm j}$ decreases and vanish as $\phi_{j} \to \phi_{J}$, consistent with previous studies~\cite{baity2017emergent,bertrand2016protocol,peyneau2008frictionless}.
The two models represent systems that consist of $N=2000$ (unless otherwise specified) bi-disperse (BD) and poly-disperse (PD) spheres,  interacting via a purely-repulsive, harmonic potential $v_{ij}(r) = \frac{1}{2} (1- r/D_{ij})^2$ (zero if $r>D_{ij}$), where $r$ is the inter-particle distance and $D_{ij} = (D_i + D_j)/2$ is the mean diameter (see Methods). 
Two independent protocols are used to create initially unjammed states whose jamming densities $\phi_j$ are above  $\phi_J$  (see Methods for details).
(i) Mechanical annealing of the BD system by the application of cyclic athermal quasistatic shear (AQS) results in unjamming of packings in the density range above $\phi_J \simeq 0.648$, as described in \cite{pallabi}.
These unjammed configurations correspond to packings with  jamming densities $\phi_{j} \in [0.648, 0.661]$. 
(ii) Thermal annealing of the PD system, with the help of an extremely efficient Monte Carlo algorithm which involves artificial swap dynamics~\cite{BCNO2016PRL}, is used to generate configurations with jamming densities $\phi_j \in [0.655, 0.69]$, above  $\phi_J \simeq 0.655$.

We first show that, an unjammed configuration at $\phi < \phi_j$, where $\phi_j > \phi_J$, can be jammed at a certain strain $\gamma_j$ 
by uniform constant volume AQS. 
The onset of shear jamming is characterised consistently by a steep increase of the shear stress $\sigma_{xz}$ (Fig. \ref{Details_of_uniform}\subref{uniform_shear_stress} and \subref{stress_PD}), of the non-rattler contact number $Z_{NR}$ (Fig. \ref{Details_of_uniform}\subref{contacts_uniform}), of the pressure (Fig.~\ref{Shear_jamming_p_PE}(a) and (b) and Fig.~\ref{fig:Details-of-uniform} of Supplementary Information (SI)),
and of the potential energy $PE$ (Fig.~\ref{Shear_jamming_p_PE}(c) and (d)),
around $\gamma_j$. We observe that $Z_{NR}$ exceeds the isostatic value $Z_{iso}= 2 D =6$, where $D=3$ is the spatial dimensionality, for $\gamma > \gamma_{j}$, indicating that the shear jammed systems are mechanically stable. 
The non-rattler contact number $Z_{NR}$ jumps discontinuously at $\gamma_j$ (Fig. \ref{Details_of_uniform}\subref{contacts_uniform}), associated with an abrupt increase of the potential energy $PE$
(Fig.~\ref{Details_of_uniform}\subref{transition}). 
The value of $\gamma_{j}$, as well as the stress overshoot amplitude,
depends on the distance to the isotropic jamming  
 $\delta \phi = \phi_{j} - \phi$,  and the value of
$\phi_{j}$ that characterizes the degree of mechanical/thermal annealing in the initial preparation procedure (Fig.~\ref{Details_of_uniform} \subref{stress_PD}). 
The data of  $PE(Z_{NR})$, 
on the other hand, follow a universal function on the jamming side $Z_{NR} > Z_{iso}$, that is independent of the jamming strain $\gamma_{j}$, the model, and the jamming protocol (shear or compression), see (Fig.~\ref{Details_of_uniform}\subref{transition}).
The data for $\phi_j \approx \phi_J$ in Fig.~\ref{Details_of_uniform} \subref{stress_PD} also clearly show that shear jamming disappears in the limit $\phi_{j} \to \phi_{J}$.


We next show that packings with $\phi_{j} > \phi_{J}$ dilate under constant pressure AQS.
For this purpose, we modify the original  AQS protocol, which is based on energy minimization at constant volume,  to minimize instead the enthalpy, allowing changes in the volume of the simulation box to ensure a fixed pressure (see Methods).
In this constant pressure AQS protocol, the system traverses only  those potential energy minima that  have 
the specified pressure, $P$.  Since the pressure is finite, the system is jammed throughout this process.  For both BD and PD models, during the constant pressure shear deformation the system dilates until reaching a {\it steady-state} at packing fraction $\phi_s$ which depends on the pressure applied (Fig.~\ref{constant_pressure}).  
Correspondingly, the stress $\sigma_{xz}$ increases initially with strain, and eventually also reaches a steady-state plateau after an overshoot (Fig.~\ref{constant_pressure}\subref{shear_stress_BD} and \ref{constant_pressure}\subref{shear_stress_PD}).
The magnitude of stress overshoot is more significant in systems with larger $\phi_j$.
The presence of a maximum at a characteristic value of the strain 
is the constant pressure analog of the maximum in stress anisotropy observed in the constant volume protocol 
as shown in the SI Fig. \ref{macro_friction_rescaled} (a) and (b).  The development of the maximum in the stress anisotropy, or in the macroscopic friction $\mu = \sigma_{\rm xz}/P$, therefore, seems to be a universal feature associated with shear jamming and dilatancy,
in both frictionless~\cite{kumar2016memory} and frictional systems, under both uniform~\cite{Sarkar2016,sarkar2017shear,seto2019shear,Zhao2019} and cyclic shear deformations~\cite{Otsuki2018}.

The amount of dilation, $\delta \phi = \phi_{init} - \phi_s$, where $\phi_{init}$ is the initial density, increases with $\phi_j$ and decreases with $P$, as seen from Fig. \ref{constant_pressure}\subref{packing_fraction_BD} and \ref{constant_pressure}\subref{packing_fraction_PD} and shown in Fig. \ref{amount_dilation} of the SI. In the limit $\phi_{j} \to \phi_{J}$ and $P\to 0$, the dilation effect disappears ($\delta \phi \to 0$), which is consistent with previous results~\cite{peyneau2008frictionless}. 
The PD model shows more significant dilation, because higher $\phi_{j}$, relative to $\phi_J$, is obtained, thanks to the efficient swap algorithm.


Steady-states, which are reached at sufficiently large strains under both constant pressure and constant volume shear deformations, are 
 memoryless: they follow equations of states (EOSs), $P_{s}(\phi_{s})$ and $\sigma_{xz,s}(\phi_{s})$,  which are independent of  initial conditions ($\phi_j$), as  
shown in Fig.~\ref{steady_state_cp}\subref{pressure_steady_EOS} and {{\ref{steady_state_cp}\subref{shear_stress_steady_EOS}}}.
Extrapolating the EOSs to the limit of zero pressure and stress, we find that the steady-states converge to a {\it critical state} at density $\phi_{c}$, i.e., $P_{s}(\phi_{s} \to \phi_{c}) \to 0$ and  $\sigma_{xz,s}(\phi_{s} \to \phi_{c}) \to 0$,
where $\phi_{c} \approx 0.648$ for the BD and  $\phi_{c} \approx 0.656$  for the PD models (Fig.~\ref{steady_state_cp}\subref{steady_state_pf_BD}). 
Within our numerical precision, the critical-state density $\phi_{c}$ coincides with the J-point density $\phi_{J}$ in large systems (see Fig. \ref{finite} in SI for finite-size analysis), $\phi_{c} \simeq \phi_{J}$, which confirms the absence of dilatancy in the limit $\phi_{j} \to \phi_{J}$.


Despite the fact that the steady-state stress is anisotropic,  $P_s(\phi_s)$ agrees well with the isotropic EOS, $P_{iso}(\phi)$,
obtained by an isotropic compression from $\phi_{J}$ (Fig.  \ref{steady_state_cp}\subref{pressure_steady_EOS}).
The critical scaling of $P_{s}$ 
also obeys a linear relationship, $P_{s}(\phi_{s}) \sim \phi_{s} - \phi_{c}$,  
as in the isotropic jamming case, where $P_{iso} \sim \phi - \phi_J$~\cite{o2003jamming}.
Fig.~\ref{steady_state_cp}\subref{pressure_steady_EOS}) further shows that, up a scale factor, the EOSs for pressure 
collapse onto the same master curve, that is not only independent of the initial condition ($\phi_{\rm j}$), but also the polydispersity (BD or PD model), and the jamming protocol (constant volume shear, constant pressure shear, or isotropic compression).
The  stress EOS $\sigma_{xz,s}(\phi_s)$ of steady-states ({Fig.\ref{steady_state_cp}\subref{shear_stress_steady_EOS}}) for the different shear protocols collapses on to a master curve, but unlike pressure, we cannot compare with the isotropic compression case, where the shear stress is always zero. 
Fig. ~\ref{steady_state_cp} (c) shows the steady-state packing fraction $\phi_{s}$ {\it vs.} pressure, indicating more clearly the approach to the asymptotic density as pressure goes to zero, independently of protocol, but different for the two studied systems. 
Figure~\ref{steady_state_cp} (d) shows that, 
the macroscopic friction of steady states $\mu_{s} = \sigma_{xz, s}/P_{s}$ is non-zero, and slowly decreases with pressure as $\mu_{s} = \mu_0 - c P_s^{\beta}$, { where $\mu_0 = 0.113, \beta = 0.453$ for the BD model, and $\mu_0 = 0.122, \beta = 0.458$ for the PD model, which are model-independent within the numerical error.}
The values of $\mu_{0}$ are also close to the previously reported data $\mu_{0} \simeq 0.1$ for mono-disperse spheres with Hertzian interactions~\cite{peyneau2008frictionless}. This scaling of $\mu_s$ suggests that, near the critical-state ($\phi_s \to \phi_c$), the stress is proportional to the pressure, $\sigma_{xz,s} \sim \mu_0  P_s$, and the stress EOS is linear,  $\sigma_{xz,s}(\phi_s) \sim \phi_s - \phi_c$, as confirmed in {SI \ref{scaling} Fig (b)}. Further details may be found in the SI, Fig.s S6 - S9. 
To summarize the above described behaviors on shear jamming and dilatancy, we propose a  generalized zero-temperature jamming phase diagram. 
The original jamming phase diagram, introduced by Liu and Nagel~\cite{liu1998nonlinear}, 
conjectures that, in the athermal limit,  the jammed states at $\phi_{J}$ should be extremely fragile under shear -- the yield stress vanishes at $\phi_{J}$ continuously from above jamming,  $\sigma_{Y}(\phi_{J}) = 0$,
suggesting that infinitesimal shear stress is required to yield (unjam) a packing at $\phi_{J}$. 
While this  picture is well supported by previous numerical studies where  $\phi_{j} \approx \phi_{J}$~\cite{heussinger2009jamming, o2003jamming, peyneau2008frictionless}, 
here we show explicitly 
a remarkable discontinuity of 
the yield stress $\sigma_{Y}$ (as well as the yield pressure $P_{Y}$) at the jamming density $\phi_j$, when $\phi_{j} > \phi_{J}$ (Fig. \ref{EOS} for the PD system; See SI Fig. \ref{Shear_jamming} for BD). This discontinuous nature  is independent  of the definition of $\sigma_{Y}$ (here we define $\sigma_{Y} = \sigma_{s}$, see {SI Fig.\ref{yield_Stress} for other definitions).  

On the contrary, the pressure $P_{iso}$ under isotropic compression vanishes continuously at $\phi_{j}$ (Fig. \ref{EOS}(a)), which is independent of $\phi_{j}$, as shown previously~\cite{chaudhuri2010jamming}. It demonstrates the reason why under constant pressure shear, the volume expands from the initial isotropic states to the final steady-states (Fig.~\ref{constant_pressure}), and under constant volume shear, the pressure increases ({ SI Fig.~\ref{Shear_jamming_p_PE} a) and b)}), {see Fig.~\ref{EOS}(a)}.
The unjammed states below $
\phi_{j}$ jam under constant volume shear, as shown  in Fig.~\ref{Details_of_uniform}.
Interestingly, the yield stress $\sigma_{Y}$ of shear jammed systems at a constant density  $\phi$ below $\phi_{j}$ appear to be a continuation of that of isotropically jammed ones. This observation is consistent with the universality of the EOSs as shown in Fig.~\ref{steady_state_cp}.
We therefore generalize the zero-temperature jamming phase diagram for frictionless spheres to arbitrary $\phi_j$,
as shown { schematically} in  Fig. \ref{EOS}, where the stress jump $\sigma_{Y}(\phi_{j})$ at  the isotropic jamming transition point $\phi_{j}$ vanishes as $\phi_{j} \to \phi_{J}$, as does  the regime of shear jamming. 
We conclude 
by firstly comparing the dilatancy effect between amorphous and lattice assemblies.
In the original paper ~\cite{reynolds1885lvii} where  the concept of ``dilatancy" was introduced for the first time, Reynolds proposed a pure geometric mechanism based on the idea that one type of lattice packing (e.g., face-centered cubic) could expand its volume under shear by transforming into another type of lattice packing (e.g., body-centered cubic).
Here we recover the same geometric mechanism for amorphous packings, which has been missed  in previous studies~\cite{peyneau2008frictionless, azema2015internal}.
Like lattices, the amorphous ensemble also includes multiple states with different packing densities,  
although jammed packings at densities above  $\phi_J$ are exponentially more abundant~\cite{parisi2010mean}.
The paths connecting these states, driven by external agitations such as shear, are accompanied by dilatancy, shear jamming, and additional rich phenomena such as avalanches, plasticity, shear  softening and hardening, and yielding. Although generic protocols lead to jammed systems with  $\phi_j \sim \phi_J$, where friction is necessary for dilatancy~\cite{Clavaud_2017,singh2018constitutive}, 
here we propose a novel approach based on cyclic shear, which  can be reproduced in experiments to generate packings with $\phi_j > \phi_J$.
Our research therefore opens the way for experimental studies on exploring the complex phase space of jamming.

\appendix
\section*{METHODS}
\noindent{\bf Models} 

\noindent (i) Bi-disperse model.
The bi-disperse system consists of $N$ equal-mass spheres with a diameter ratio $D_1/D_2 = 1.4$ and a number ratio $N_1/N_2 = 1$.  

\noindent (ii) Poly-disperse model.
The PD system contains  $N$ equal-mass spheres whose diameter distribution is characterized by $P(D) \sim D^{-3}$, for $D_{\rm min} \leq D \leq D_{\rm min}/0.45$.   
The volume fraction is $\phi=\rho (1/6)\pi \overline{D^3}$, where $\rho$ is the number density $N/V$, and $V$ is the volume of simulation box. 

\noindent{\bf Constant volume athermal quasi-static shear}

\noindent (i) In the BD model, constant volume AQS simulations  are carried out 
using LAMMPS\cite{plimpton1995fast}. 
To simulate a uniform simple shear deformation, at each step an affine transformation is applied to the position of each particle, $ x'  =x+\delta\gamma\times z, y' =y, z'  =z$, where $\delta\gamma = 10^{-4}$. 
followed by energy minimization using the conjugate gradient (CG) method.
The CG procedure stops when the maximum 
component of the force vector is less than $10^{-16}$. The energy minimization 
stops when the maximum distance moved by any particle is less than the machine precision during 
an iteration. 
The norm of the equilibrium net force vector is of the order of  $10^{-13}$ and the maximum component  is of the order of $10^{-14}$ at the termination of minimization.

\noindent (ii) In the PD model, the affine  transformation is applied with step size $\delta \gamma = 10^{-4}$, followed by energy minimization using the FIRE algorithm~\cite{bitzek2006structural}. The minimization procedure stops when the fraction of force balanced particles with net force magnitude  $|f| \leq 10^{-14}$ 
grows above 0.995.

\noindent{\bf Constant pressure athermal quasi-static shear}

In constant pressure AQS simulations,  the energy minimization is replaced by the minimization of enthalpy $H = U + PV$
at the imposed pressure $P$. (i) In the BD model, the minimization stops when the maximum distance moved 
by any particle during a minimization step is less than the machine precision. (ii) In the PD model, the minimization stops if the fraction of force balanced particles 
is greater than 0.995, and the deviation from the target pressure is less than $10^{-4}$.

\noindent{\bf Protocols to prepare initial configurations} 

\noindent (i) Mechanical annealing by cyclic athermal quasistatic shear for the BD model.
We first use the method in \cite{chaudhuri2010jamming} to generate packings with jamming density $\phi_J\approx0.648$.
The initial configurations are hard-sphere configurations at a packing fraction of $\phi=0.363$, which are equilibrated using the Monte-Carlo (MC) algorithm.
{We switch to the harmonic soft-sphere potential, rapidly compress the configurations by rescaling the volume of the simulation box (till  $\beta P/\rho$ decays to $\sim 1000$, where $\beta$ is the inverse temperature), and remove the resulting overlaps by using MC simulations.}  
{The temperature is then switched off, and the system is further quasi-statically compressed, by inflating the particles uniformly, followed by energy minimization using the conjugate gradient method.} 
The compression stops when the energy per particle $e=E/N$, after minimization, remains above $10^{-16}$. This is used as the criterion for jamming. 
Then the system is slowly decompressed till $e< 10^{-16}$, which generates configurations corresponding to jamming density $\phi_J\approx0.648$.

We then use mechanical annealing to increase the jamming density from $\phi_J$ to $\phi_j > \phi_J$.
The configurations obtained from the above procedure are  compressed to various over-jamming densities $\phi > \phi_J$, and are unjammed 
using cyclic AQS, $\gamma = 0\rightarrow\gamma_{max}\rightarrow0\rightarrow -\gamma_{max}\rightarrow0$, where 
 the strain amplitude $\gamma_{max}=0.07$ \cite{pallabi}, and the strain step { $\delta \gamma = 10^{-3}$}. These configurations correspond to jamming densities $\phi_j > \phi_J$. {See SI Fig.~\ref{dependence_of_jamming_density} for the dependence of $\phi_j$ on protocol parameters.} 

\noindent (ii)  Thermal annealing by a swap algorithm  for the PD model.   
We first prepare dense equilibrium HS configurations at  $\phi_{g}$, using the  the swap algorithm~\cite{BCNO2016PRL}. At each swap MC step, we exchange the positions of two randomly picked particles as long as they do not overlap with other particles. Combined with standard event-driven molecular dynamics (MD), such non-local swap moves significantly speed up the equilibration procedure. 
The poly-dispersity of the model suppresses crystallization even in deep annealing, and optimizes the efficiency  of the algorithm~\cite{BCNO2016PRL}.

For each equilibrium configuration at $\phi_{g}$, we then perform a rapid quench to generate the jammed configuration at $\phi_{j}$ (see Ref.~\cite{berthier2016growing} for the relationship between $\phi_{g}$ and $\phi_j$). In particular, the J-point state at $\phi_J \simeq 0.655$ are quenched from random initial configurations with $\phi_{g} = 0$~\cite{o2003jamming}.
The rapid quench is realized by 
inflating the particle sizes instantaneously to reach the target density, switching to the harmonic soft-sphere  potential, and then minimizing the total potential energy using the FIRE algorithm~\cite{bitzek2006structural}. The same jamming criterion is used as in the BD model.

\noindent{\bf Calculation of the stress tensor and the pressure} 
 
The stress tensor is calculated using the formula,
\begin{equation}
	\hat{\sigma}=-\frac{1}{V} \sum_{i<j} \vec{f}_{ij} \otimes \vec{r}_{ij},
\end{equation}
where $\vec{f}_{ij}$ and $\vec{r}_{ij}$ are the inter-particle force and distance vectors between particles $i$ and $j$.
The pressure $P$ is related to the trace of the stress tensor, $P = - (\sigma_{xx} + \sigma_{yy} + \sigma_{zz})/3$, which can be written as,
\begin{equation}
	P =\frac{1}{3V} \sum_{i<j} \vec{f}_{ij} \cdot \vec{r}_{ij}.
\end{equation}

\bibliographystyle{unsrt}
\bibliography{dilatancy}

\begin{thebibliography}{10}

\bibitem{reynolds1885lvii}
Osborne Reynolds.
\newblock Lvii. on the dilatancy of media composed of rigid particles in
  contact. with experimental illustrations.
\newblock {\em The London, Edinburgh, and Dublin Philosophical Magazine and
  Journal of Science}, 20(127):469--481, 1885.

\bibitem{peyneau2008frictionless}
Pierre-Emmanuel Peyneau and Jean-No{\"e}l Roux.
\newblock Frictionless bead packs have macroscopic friction, but no dilatancy.
\newblock {\em Physical review E}, 78(1):011307, 2008.

\bibitem{baity2017emergent}
Marco Baity-Jesi, Carl~P Goodrich, Andrea~J Liu, Sidney~R Nagel, and James~P
  Sethna.
\newblock Emergent so(3) symmetry of the frictionless shear jamming transition.
\newblock {\em Journal of Statistical Physics}, 167(3-4):735--748, 2017.

\bibitem{liu1998nonlinear}
Andrea~J Liu and Sidney~R Nagel.
\newblock Nonlinear dynamics: Jamming is not just cool any more.
\newblock {\em Nature}, 396(6706):21, 1998.

\bibitem{o2003jamming}
Corey~S O'hern, Leonardo~E Silbert, Andrea~J Liu, and Sidney~R Nagel.
\newblock Jamming at zero temperature and zero applied stress: The epitome of
  disorder.
\newblock {\em Physical Review E}, 68(1):011306, 2003.

\bibitem{chaudhuri2010jamming}
Pinaki Chaudhuri, Ludovic Berthier, and Srikanth Sastry.
\newblock Jamming transitions in amorphous packings of frictionless spheres
  occur over a continuous range of volume fractions.
\newblock {\em Physical review letters}, 104(16):165701, 2010.

\bibitem{heussinger2009jamming}
Claus Heussinger and Jean-Louis Barrat.
\newblock Jamming transition as probed by quasistatic shear flow.
\newblock {\em Physical review letters}, 102(21):218303, 2009.

\bibitem{Cates_1998}
M~E Cates, J~P Wittmer, J.-P. Bouchaud, and P~Claudin.
\newblock {Jamming, force chains, and fragile matter}.
\newblock {\em Phys. Rev. Lett.}, 81:1841--1844, 1998.

\bibitem{parisi2010mean}
Giorgio Parisi and Francesco Zamponi.
\newblock Mean-field theory of hard sphere glasses and jamming.
\newblock {\em Reviews of Modern Physics}, 82(1):789, 2010.

\bibitem{ozawa2012jamming}
Misaki Ozawa, Takeshi Kuroiwa, Atsushi Ikeda, and Kunimasa Miyazaki.
\newblock Jamming transition and inherent structures of hard spheres and disks.
\newblock {\em Physical review letters}, 109(20):205701, 2012.

\bibitem{bi2011jamming}
Dapeng Bi, Jie Zhang, Bulbul Chakraborty, and Robert~P Behringer.
\newblock Jamming by shear.
\newblock {\em Nature}, 480(7377):355, 2011.

\bibitem{ren2013reynolds}
Jie Ren, Joshua~A Dijksman, and Robert~P Behringer.
\newblock Reynolds pressure and relaxation in a sheared granular system.
\newblock {\em Physical review letters}, 110(1):018302, 2013.

\bibitem{sarkar2013origin}
Sumantra Sarkar, Dapeng Bi, Jie Zhang, RP~Behringer, and Bulbul Chakraborty.
\newblock Origin of rigidity in dry granular solids.
\newblock {\em Phys. Rev. Lett.}, 111(6):068301, 2013.

\bibitem{Sarkar2016}
S.~Sarkar, D.~Bi, J.~Zhang, J.~Ren, R.P. Behringer, and B.~Chakraborty.
\newblock {Shear-induced rigidity of frictional particles: Analysis of emergent
  order in stress space}.
\newblock {\em Physical Review E}, 93(4), 2016.

\bibitem{kumar2016memory}
Nishant Kumar and Stefan Luding.
\newblock Memory of jamming--multiscale models for soft and granular matter.
\newblock {\em Granular Matter}, 18(3):58, 2016.

\bibitem{vinutha2016disentangling}
HA~Vinutha and Srikanth Sastry.
\newblock Disentangling the role of structure and friction in shear jamming.
\newblock {\em Nature Physics}, 12(6):578, 2016.

\bibitem{vinutha_kabir}
H~A Vinutha, Kabir Ramola, Bulbul Chakraborty, and Srikanth Sastry.
\newblock Timescale divergence at the shear jamming transition.
\newblock {\em arXiv preprint arXiv:1903.01496}, 2019.

\bibitem{Setoa}
Ryohei Seto, Abhinendra Singh, Bulbul Chakraborty, Morton~M Denn, and Jeffrey~F
  Morris.
\newblock {Shear jamming and fragility in dense suspensions}.

\bibitem{Otsuki2017}
Michio Otsuki and Hisao Hayakawa.
\newblock {Discontinuous change of shear modulus for frictional jammed granular
  materials}.
\newblock {\em Physical Review E}, 95(6):1--10, 2017.

\bibitem{Otsuki2018}
Michio Otsuki and Hisao Hayakawa.
\newblock Shear jamming, discontinuous shear thickening, and fragile states in
  dry granular materials under oscillatory shear.
\newblock {\em Physical Review E}, 101(3):032905, 2020.

\bibitem{Jineaat6387}
Yuliang Jin, Pierfrancesco Urbani, Francesco Zamponi, and Hajime Yoshino.
\newblock A stability-reversibility map unifies elasticity, plasticity,
  yielding, and jamming in hard sphere glasses.
\newblock {\em Science Advances}, 4(12), 2018.

\bibitem{Zhao2019}
Yiqiu Zhao, Jonathan Bar{\'{e}}s, Hu~Zheng, Joshua E.~S. Socolar, and Robert~P.
  Behringer.
\newblock {Shear-Jammed, Fragile, and Steady States in Homogeneously Strained
  Granular Materials}.
\newblock {\em Physical Review Letters}, 123(15):1--7, 2019.

\bibitem{Brown_2012}
E.~Brown and H.~M. Jaeger.
\newblock The role of dilation and confining stresses in shear thickening of
  dense suspensions.
\newblock {\em J. Rheol.}, 56:875--923, 2012.

\bibitem{Clavaud_2017}
C{\'e}cile Clavaud, Antoine B{\'e}rut, Bloen Metzger, and Yo{\"e}l Forterre.
\newblock Revealing the frictional transition in shear-thickening suspensions.
\newblock {\em Proc. Natl. Acad. Sci. U.S.A.}, pages 5147--5152, 2017.

\bibitem{kruyt2016micromechanical}
Nicolaas~P Kruyt and L~Rothenburg.
\newblock A micromechanical study of dilatancy of granular materials.
\newblock {\em Journal of the Mechanics and Physics of Solids}, 95:411--427,
  2016.

\bibitem{rao2008introduction}
K~Kesava Rao and Prabhu~R Nott.
\newblock {\em An introduction to granular flow/K. Kesava Rao, Prabhu R. Nott.}
\newblock 2008.

\bibitem{rowe1962stress}
Peter~W Rowe.
\newblock The stress-dilatancy relation for static equilibrium of an assembly
  of particles in contact.
\newblock {\em Proceedings of the Royal Society of London. Series A.
  Mathematical and Physical Sciences}, 269(1339):500--527, 1962.

\bibitem{behringer2018physics}
Robert~P Behringer and Bulbul Chakraborty.
\newblock The physics of jamming for granular materials: a review.
\newblock {\em Reports on Progress in Physics}, 82(1):012601, 2018.

\bibitem{azema2015internal}
{\'E}milien Az{\'e}ma, Farhang Radja{\"\i}, and Jean-No{\"e}l Roux.
\newblock Internal friction and absence of dilatancy of packings of
  frictionless polygons.
\newblock {\em Physical Review E}, 91(1):010202, 2015.

\bibitem{bertrand2016protocol}
Thibault Bertrand, Robert~P Behringer, Bulbul Chakraborty, Corey~S O'Hern, and
  Mark~D Shattuck.
\newblock Protocol dependence of the jamming transition.
\newblock {\em Physical Review E}, 93(1):012901, 2016.

\bibitem{pallabi}
Pallabi Das, H~A Vinutha, and Srikanth Sastry.
\newblock Unified phase diagram of reversible-irreversible, jamming and
  yielding transitions in cyclically sheared soft sphere packings.
\newblock {\em arXiv preprint arXiv:1907.08503}, 2019.

\bibitem{BCNO2016PRL}
Ludovic Berthier, Daniele Coslovich, Andrea Ninarello, and Misaki Ozawa.
\newblock Equilibrium sampling of hard spheres up to the jamming density and
  beyond.
\newblock {\em Phys. Rev. Lett.}, 116:238002, Jun 2016.

\bibitem{sarkar2017shear}
Sumantra Sarkar, Elan Shatoff, Kabir Ramola, Romain Mari, Jeffrey Morris, and
  Bulbul Chakraborty.
\newblock Shear-induced organization of forces in dense suspensions: signatures
  of discontinuous shear thickening.
\newblock In {\em EPJ Web of Conferences}, volume 140, page 09045. EDP
  Sciences, 2017.

\bibitem{seto2019shear}
R.~Seto, A.~Singh, B.~Chakraborty, M.~M Denn, and J.~F Morris.
\newblock Shear jamming and fragility in dense suspensions.
\newblock {\em arXiv preprint arXiv:1902.04361}, 2019.

\bibitem{singh2018constitutive}
A.~Singh, R.~Mari, M.~M. Denn, and J.~F. Morris.
\newblock A constitutive model for simple shear of dense frictional
  suspensions.
\newblock {\em J. Rheol.}, 62(2):457--468, 2018.

\bibitem{plimpton1995fast}
Steve Plimpton.
\newblock Fast parallel algorithms for short-range molecular dynamics.
\newblock {\em Journal of computational physics}, 117(1):1--19, 1995.

\bibitem{bitzek2006structural}
Erik Bitzek, Pekka Koskinen, Franz G{\"a}hler, Michael Moseler, and Peter
  Gumbsch.
\newblock Structural relaxation made simple.
\newblock {\em Physical review letters}, 97(17):170201, 2006.

\bibitem{berthier2016growing}
Ludovic Berthier, Patrick Charbonneau, Yuliang Jin, Giorgio Parisi, Beatriz
  Seoane, and Francesco Zamponi.
\newblock Growing timescales and lengthscales characterizing vibrations of
  amorphous solids.
\newblock {\em Proceedings of the National Academy of Sciences},
  113(30):8397--8401, 2016.

\bibitem{coslovich2017exploring}
Daniele Coslovich, Ludovic Berthier, and Misaki Ozawa.
\newblock Exploring the jamming transition over a wide range of critical
  densities.
\newblock {\em SciPost Physics}, 3(4):027, 2017.

\bibitem{heussinger2010fluctuations}
Claus Heussinger, Pinaki Chaudhuri, and Jean-Louis Barrat.
\newblock Fluctuations and correlations during the shear flow of elastic
  particles near the jamming transition.
\newblock {\em Soft matter}, 6(13):3050--3058, 2010.

\bibitem{zheng2018jamming}
Wen Zheng, Shiyun Zhang, and Ning Xu.
\newblock Jamming of packings of frictionless particles with and without shear.
\newblock {\em Chinese Physics B}, 27(6):066102, 2018.

\bibitem{vaagberg2011glassiness}
Daniel V{\aa}gberg, Peter Olsson, and Stephen Teitel.
\newblock Glassiness, rigidity, and jamming of frictionless soft core disks.
\newblock {\em Physical Review E}, 83(3):031307, 2011.

\end{thebibliography}
\begin{acknowledgments}
  We warmly thank K.~Miyazaki, T.~Kawasaki, M.~Otsuki,  for discussions.
  Y.J. acknowledges funding from Project 11974361,  Project 11935002, and Project 11947302 supported by NSFC, from Key Research Program of Frontier  Sciences, CAS, Grant NO. ZDBS-LY-7017, and from the CAS Pioneer Hundred Talents Program. BC acknowledges support from NSF-CBET-1916877, BSF-2016188. and a Simons Fellowship in Theoretical Physics.  SS acknowledges support through the J. C. Bose Fellowship, SERB, DST, India.  BC and SS acknowledge the support of the Indo-US Virtual Networked Joint Center project titled “Emergence and Re-modeling of force chains in soft and Biological Matter No. IUSSTF/JC-026/2016.
  The computations were performed using resources at TUS-CMS, JNCASR, the HPC Cluster of ITP-CAS,  Tianhe-2 Supercomputer and National Supercomputer Center in Guangzhou.
\end{acknowledgments}

\begin{figure*}[htp]
	\subfloat[]{\includegraphics[scale=0.34]{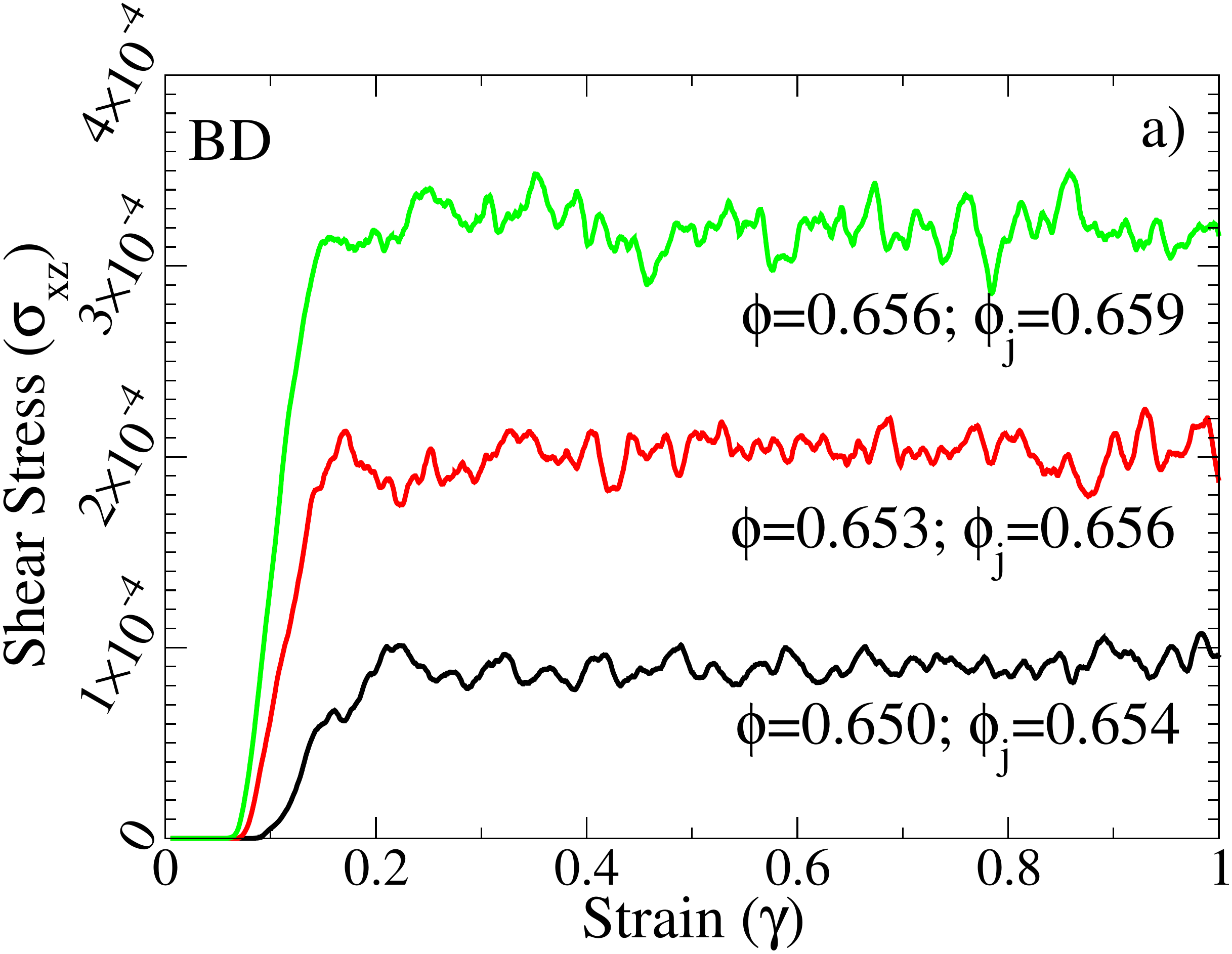}\label{uniform_shear_stress}}
	\hfill
	\subfloat[]{\includegraphics[scale=0.34]{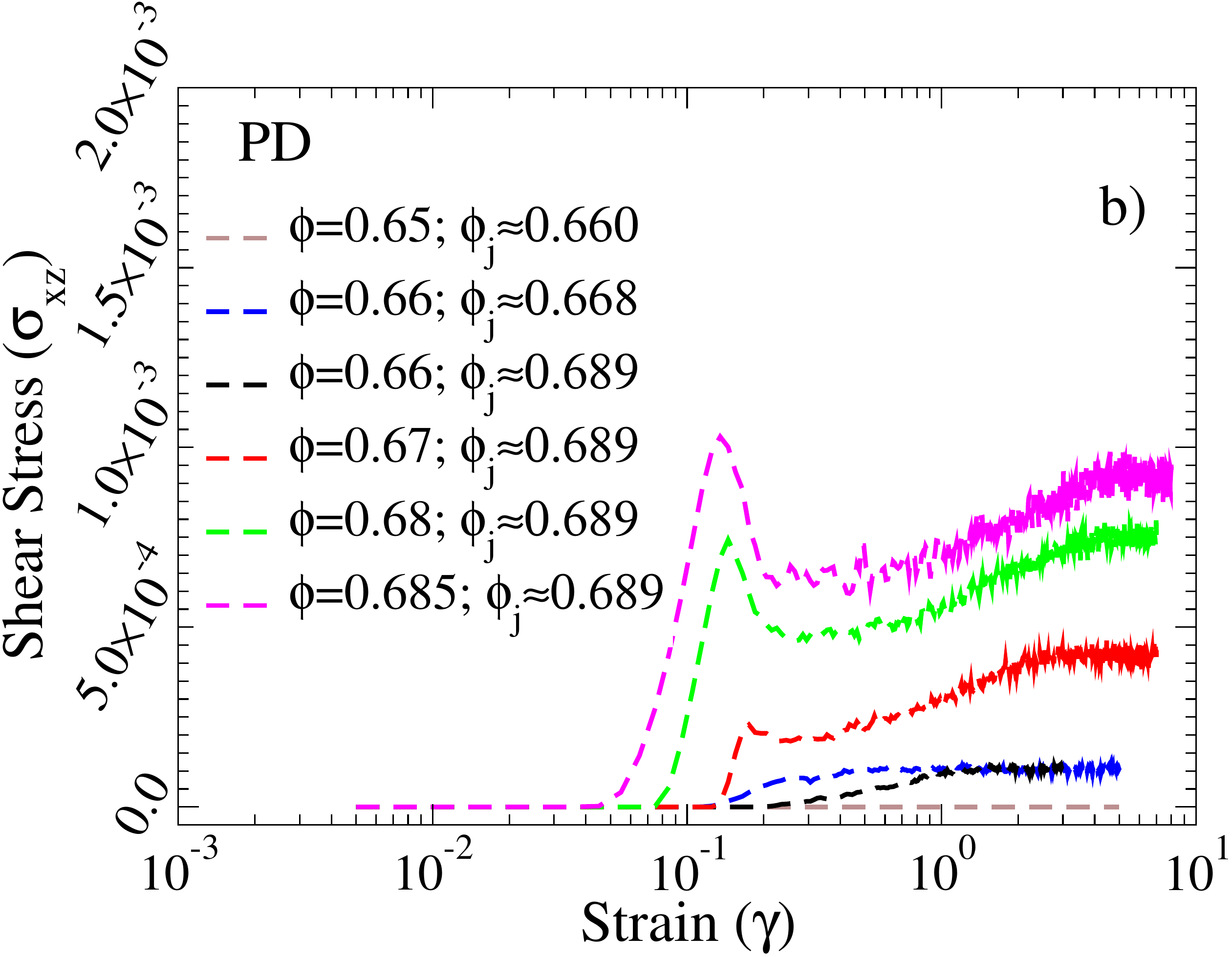}\label{stress_PD}}
	\vfill{}
	\subfloat[]{\includegraphics[scale=0.34]{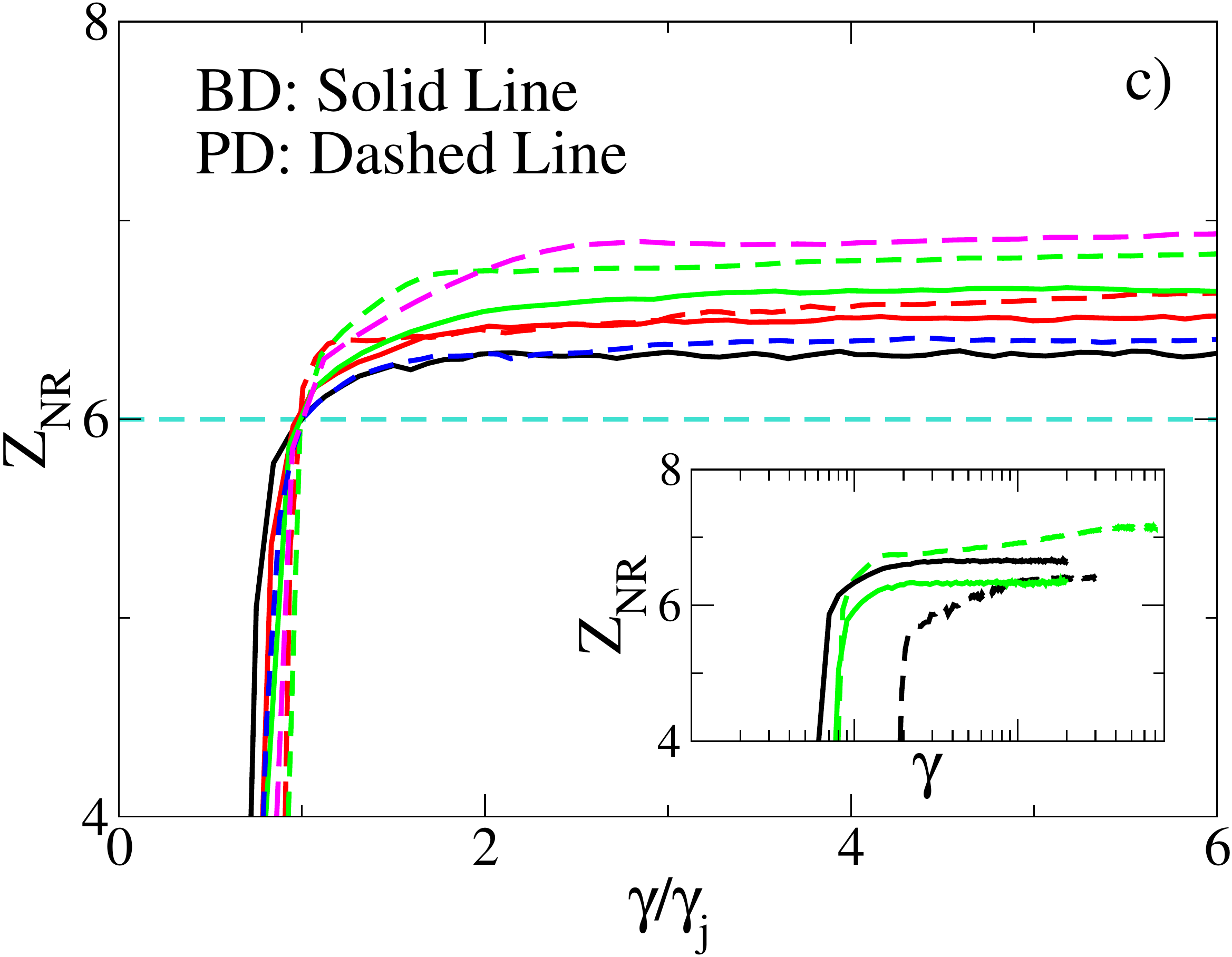}\label{contacts_uniform}}
	\hfill
	\subfloat[]{\includegraphics[scale=0.34]{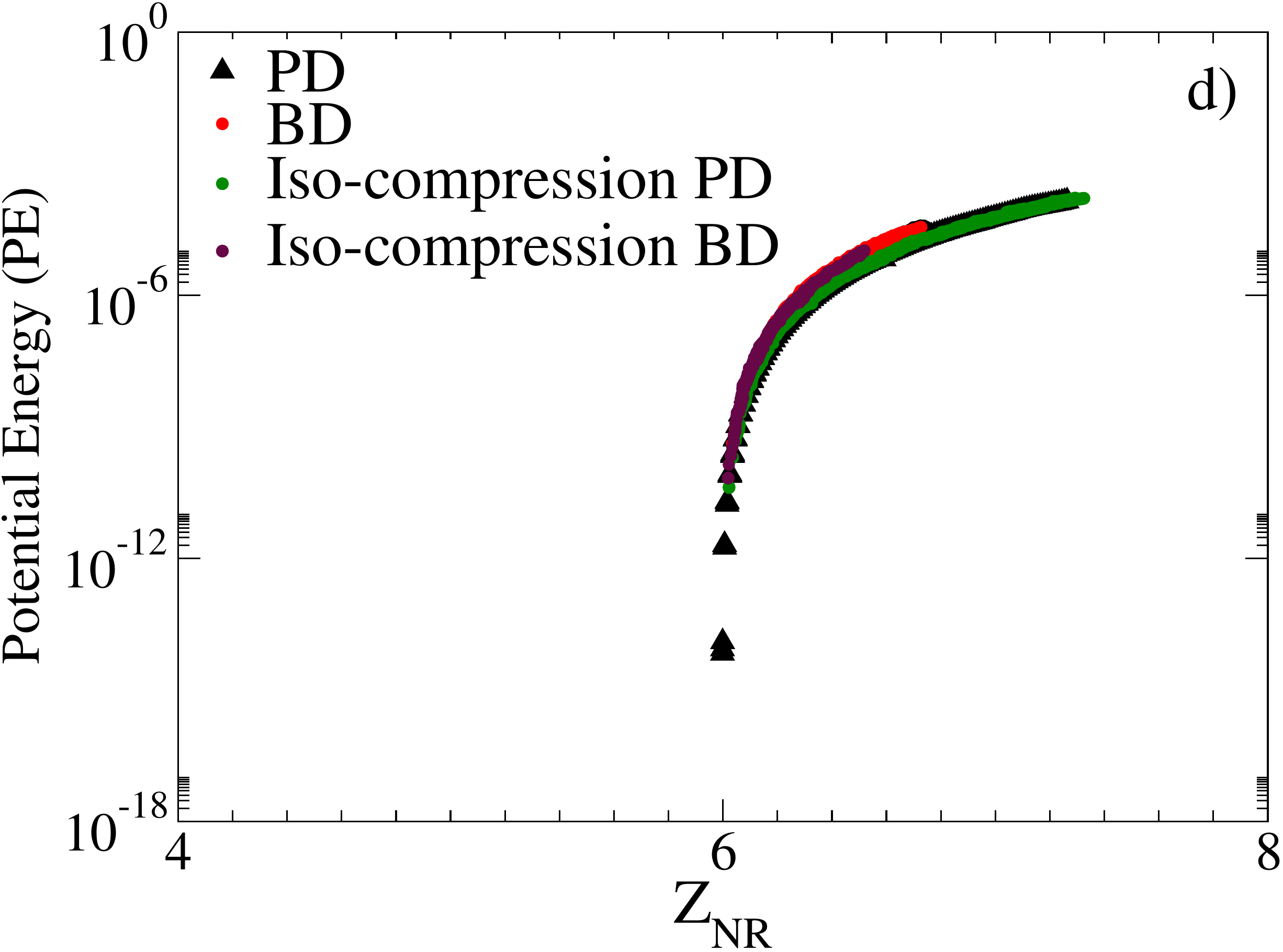}\label{transition}}	
	\caption{\label{Details_of_uniform} 
   \textbf{Shear jamming.}  
   Shear stress $\sigma_{xz}$ as a function of strain $\gamma$ for  (a) BD model, (b) PD model, and a few different $\phi_{j}$ and $\phi$. 
   (c)  Non-rattler contact number $Z_{NR}$, which is calculated after removing rattlers (particles with less than $D+1$ contacts) recursively, as a function of $\gamma/\gamma_{j}$. 
   Inset shows unscaled data, configurations at different densities jam at different strains.
   (d) The potential energy $PE$ is a universal function of  $Z_{NR}$ above jamming, for both BD and PD systems, for different $\phi_{j}$ and $\phi$ (and therefore different $\gamma_{j}$), and for both compression and shear jamming. 
   The data are averaged over 20 and 64 independent samples in  BD and PD  systems respectively. 
	}
\end{figure*}

\begin{figure*}[htp]
	\subfloat[]{\includegraphics[scale=0.35]{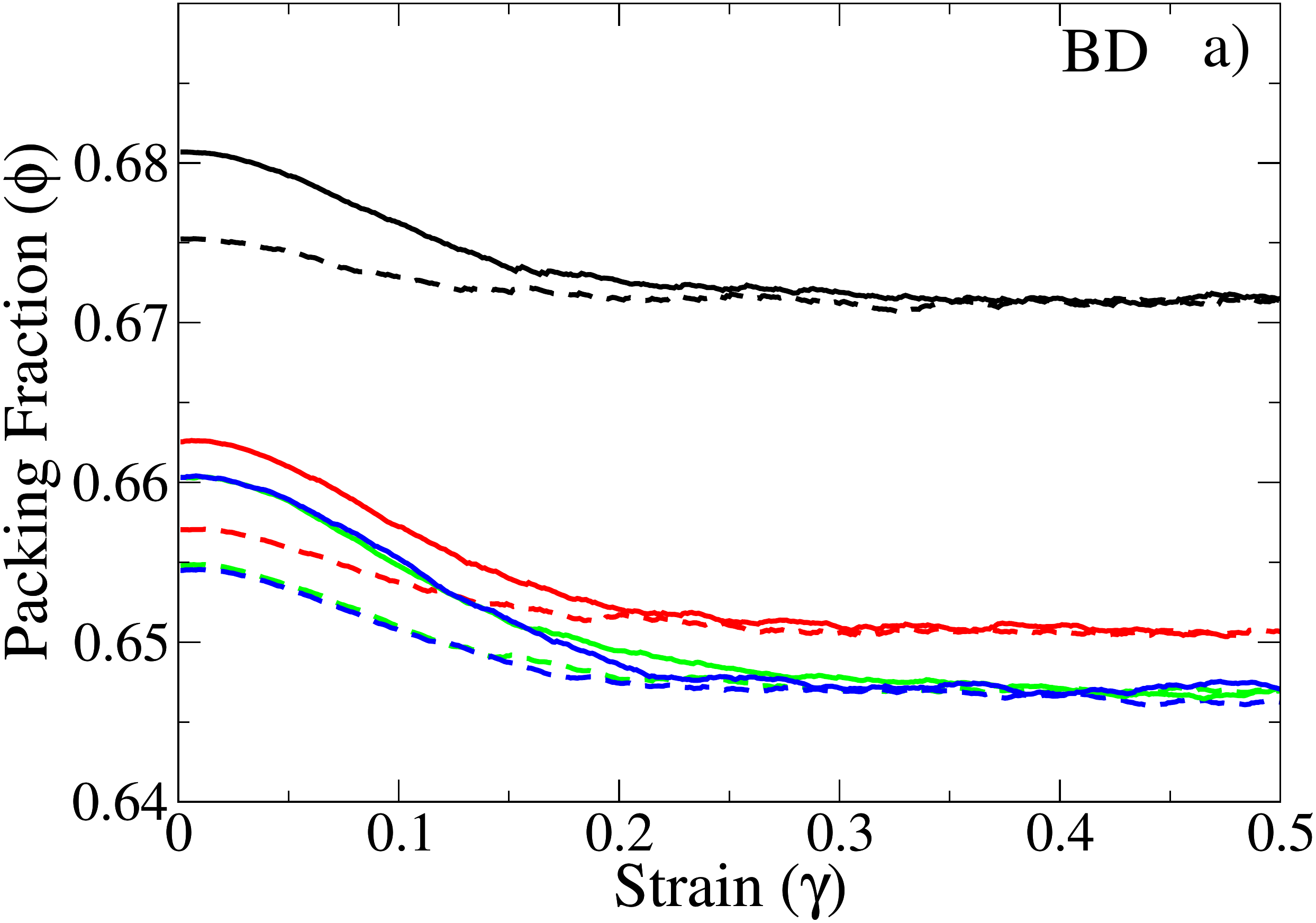}{\label{packing_fraction_BD}}}
	\hfill
	\subfloat[]{\includegraphics[scale=0.35]{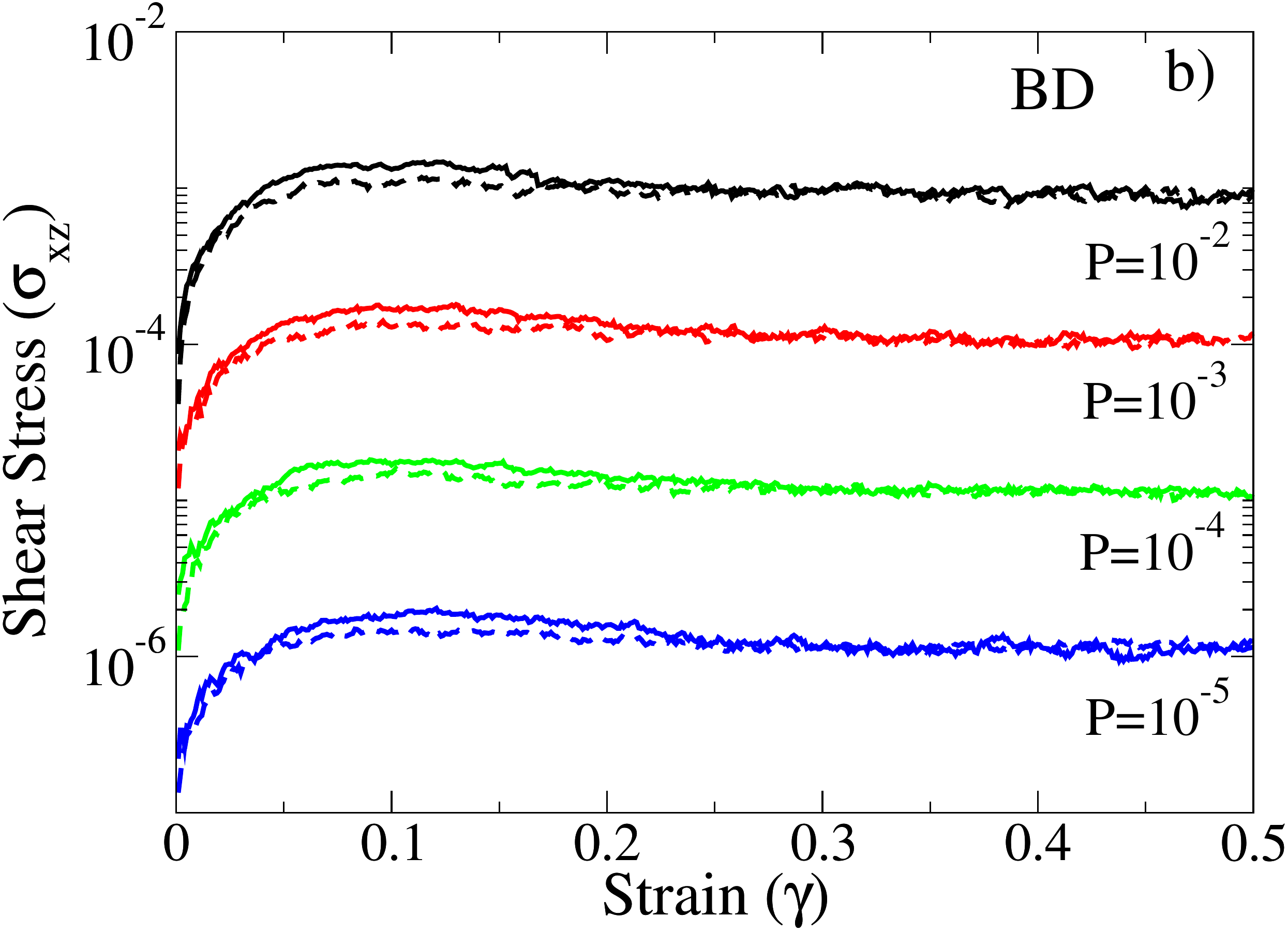}{\label{shear_stress_BD}}}
	\vfill{}
	\subfloat[]{\includegraphics[scale=0.35]{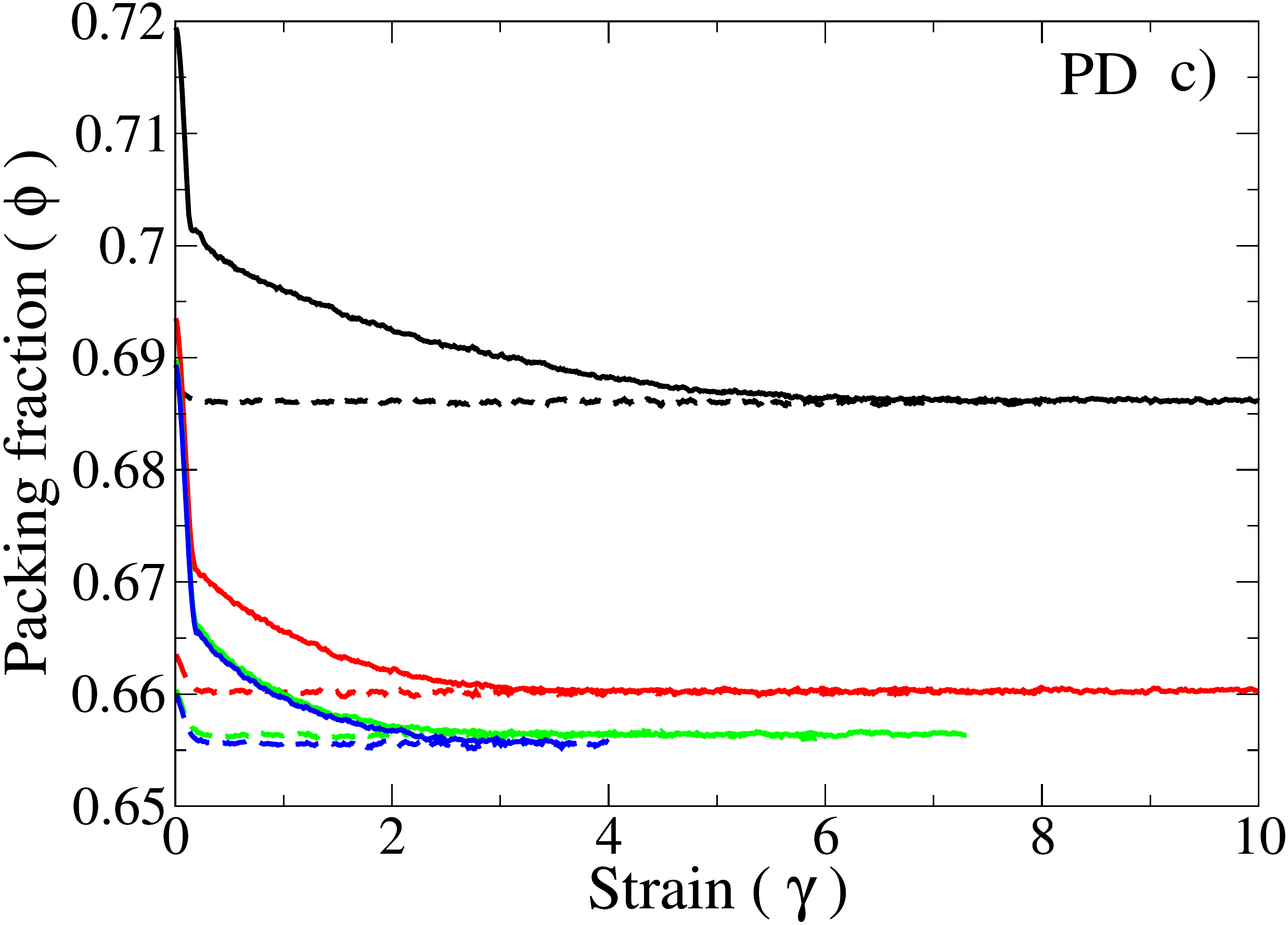}{\label{packing_fraction_PD}}}
	\hfill
	\subfloat[]{\includegraphics[scale=0.35]{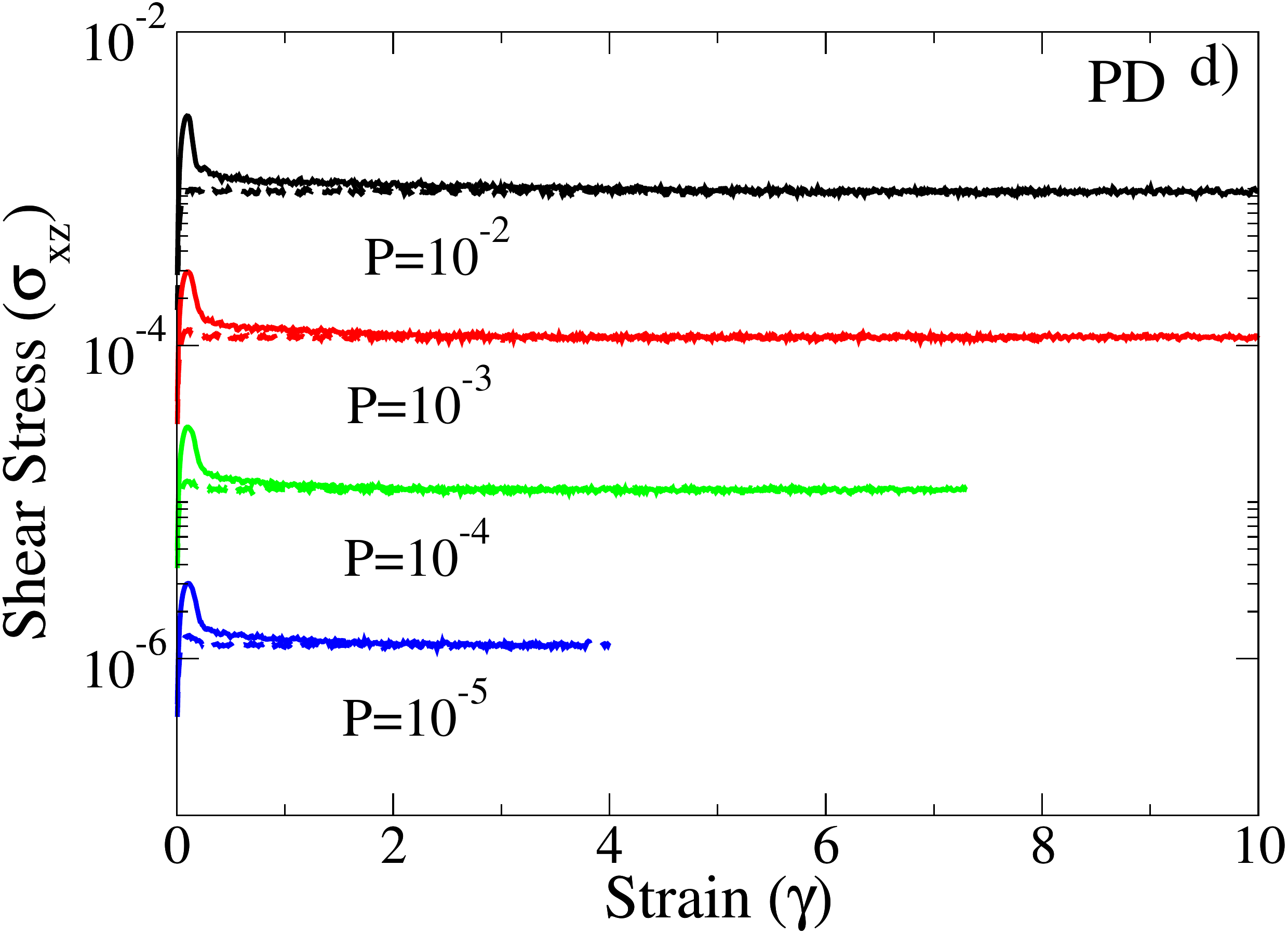}{\label{shear_stress_PD}}}
	\caption{ 
	\small{
	\label{constant_pressure} 
	\textbf{Dilatancy.} 
	The evolution of (a) packing fraction $\phi$ and (b) shear stress $\sigma_{xz}$ as functions of strain $\gamma$ under constant pressure AQS in the BD model, for $\phi_{j} = 0.654$ (dashed) and  $\phi_{j} = 0.66$ (solid), 
	and for a few different pressures $P$ (indicated in (b)). (c,d) Same data in the PD model ($P$ values indicated in (d)), for $\phi_{j} = 0.660$ (dashed) and $\phi_{j} = 0.689$ (solid). 
	 { The data are averaged over 10 and 64 independent samples in  BD and PD  systems respectively.}
	}
}
	
\end{figure*}

\begin{figure*}[htp]
        \subfloat[]{\includegraphics[scale=0.34]{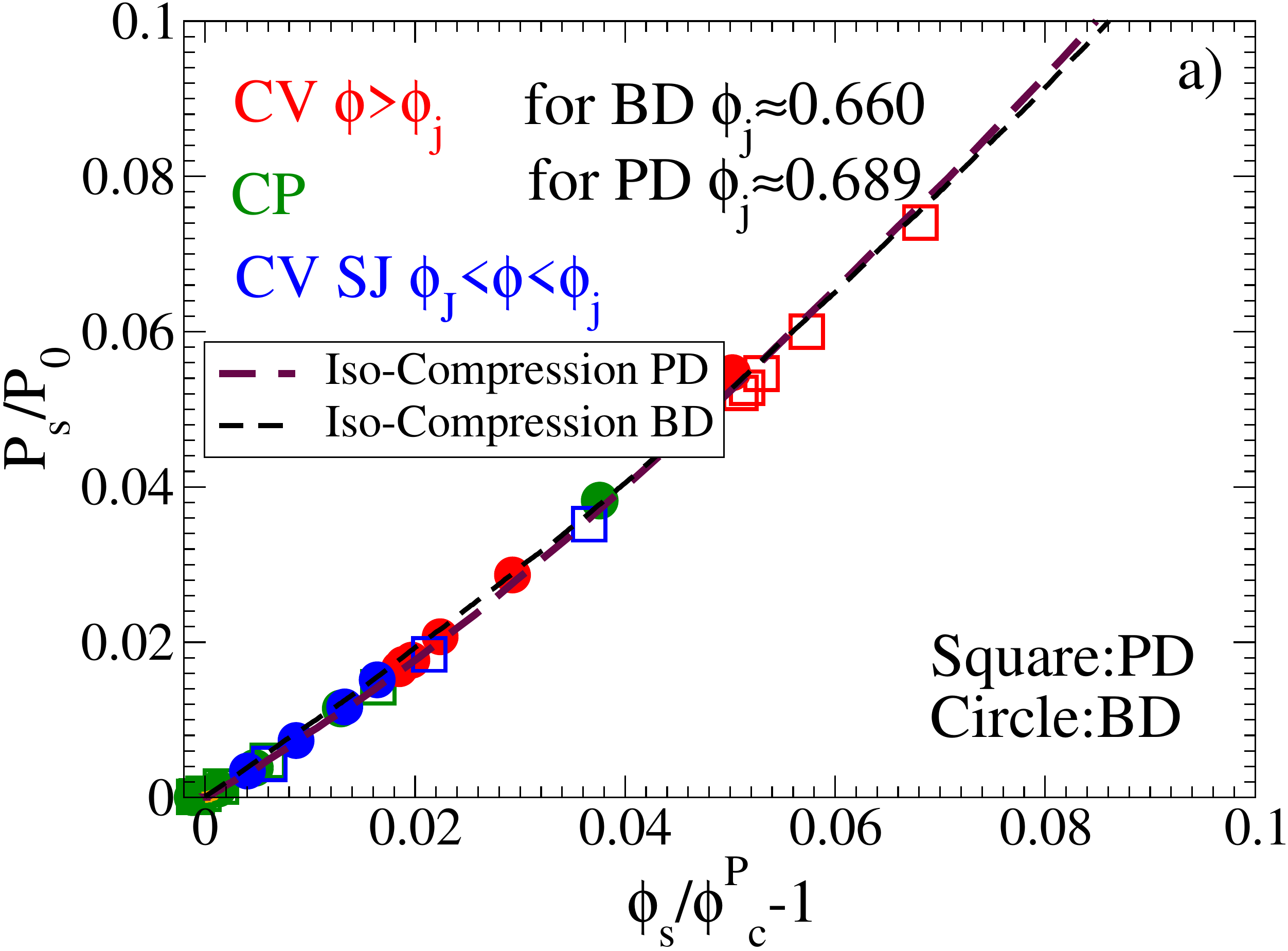}{\label{pressure_steady_EOS}}}
        \hfill
        \subfloat[]{\includegraphics[scale=0.34]{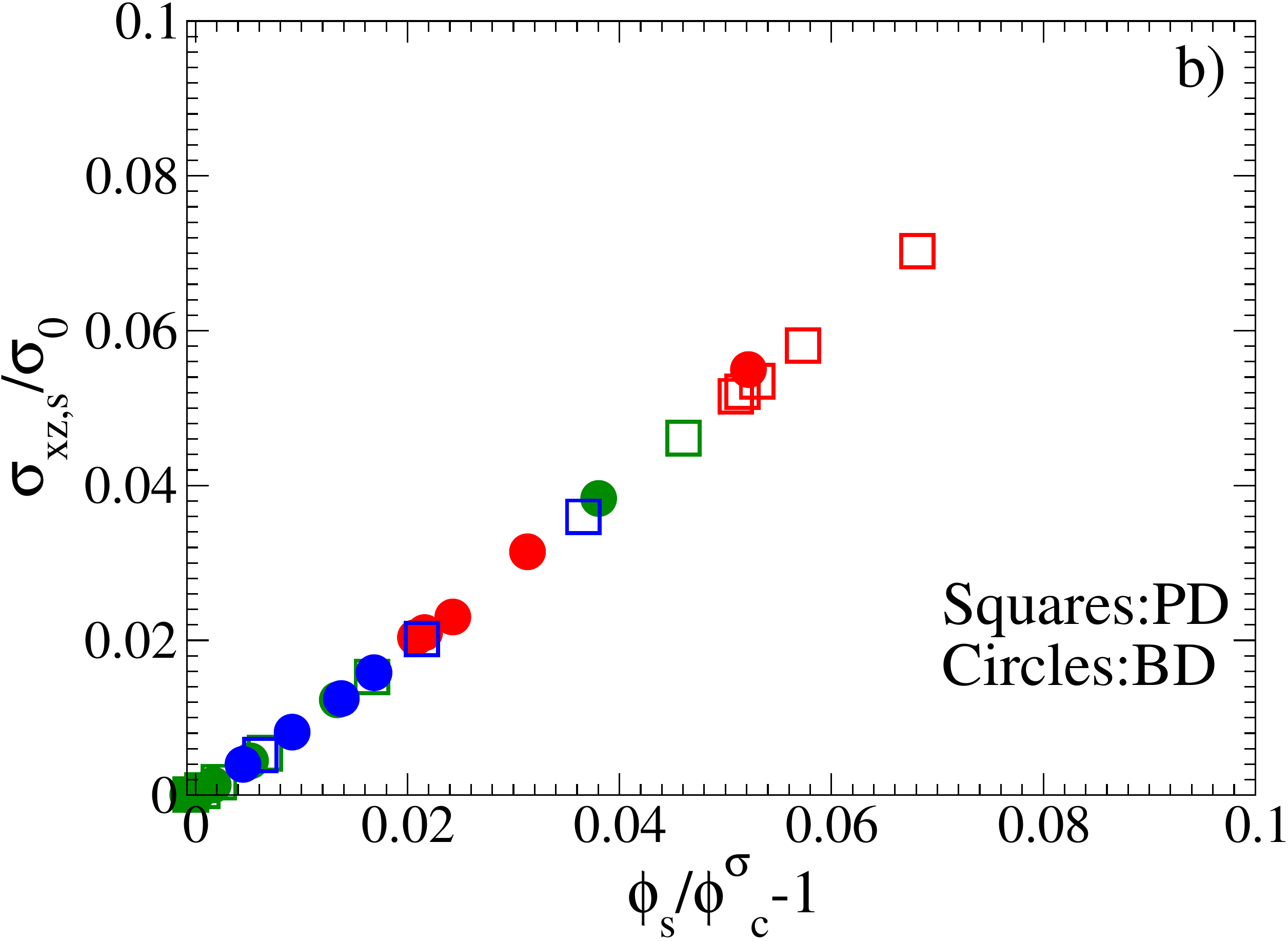}{\label{shear_stress_steady_EOS}}}
        \vfill{}
        \subfloat[]{\includegraphics[scale=0.34]{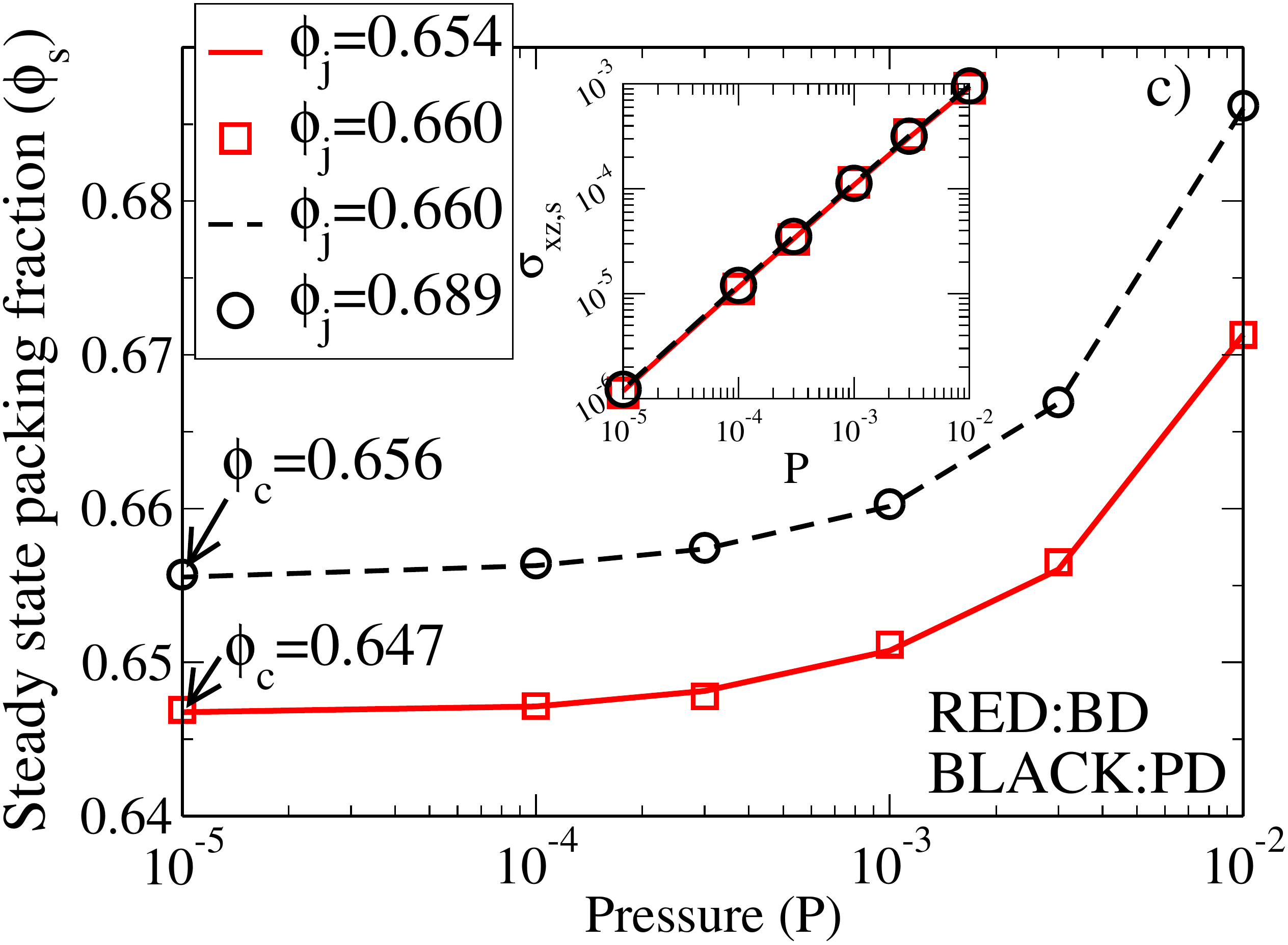}{\label{steady_state_pf_BD}}}
       	\hfill{}
        \subfloat[]{\includegraphics[scale=0.34]{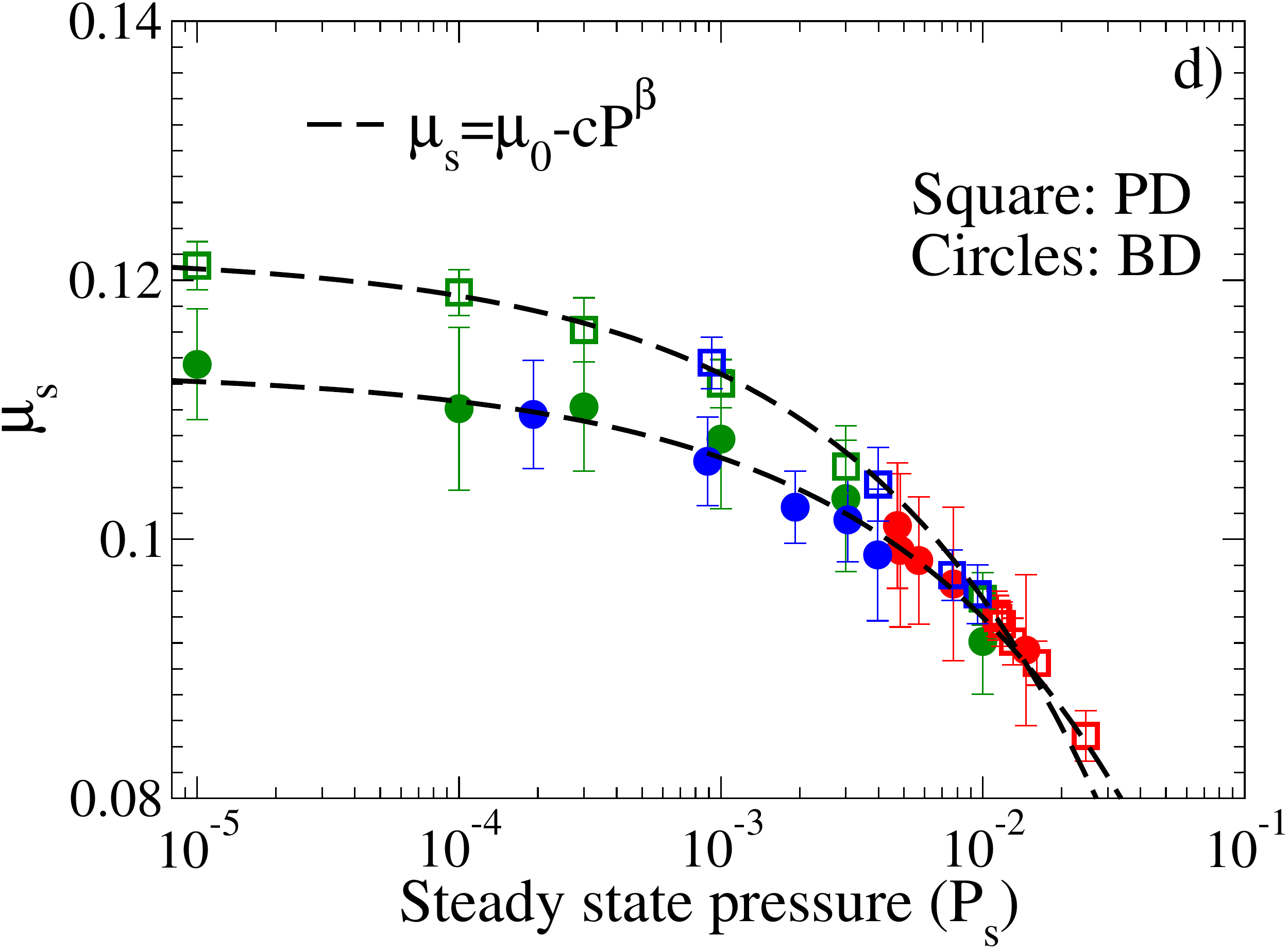}{\label{macro_friction}}}
	\caption{
        \label{steady_state_cp}\textbf{Steady-state EOSs.}
	\textbf{(a)} The steady-state pressure $P_s$ is a universal function of $\phi_s/{\phi_c}^P - 1$ after rescaling, {where for the BD model, the scaling factor $P_0 = 0.261$ and the jamming density $\phi_j = 0.66$, and for the PD model, $P_0 = 0.217$ and $\phi_j = 0.689$.}
	The data are obtained from constant volume  shear above $\phi_j$ (red), constant volume shear below $\phi_j$ where shear jamming occurs (blue), constant pressure shear (green), and isotropic compression from $\phi_J$ where $P_{iso}$ is plotted as a function of $\phi/\phi_J-1$ with {$P_0' = 0.21$ for PD and $P_0'=0.29$ for BD}.
	\textbf{(b)} {{Steady-state stress $\sigma_{xz,s}$ is a universal function of $\phi_s/{\phi_c}^\sigma-1$.
	The rescaling factor $\sigma_0=0.024$ for BD and $\sigma_0=0.021$ for PD.}}
	\textbf{(c)} Steady-state density $\phi_s$ as a function of pressure $P_s$ for different $\phi_j$, obtained from constant pressure shear. Inset: Steady-state  stress is independent of the jamming density for constant pressure shear deformation.  
	\textbf{d)} Macroscopic friction $\mu_s$ of steady-states as a function of pressure $P_{s}$. 
	}
\end{figure*}

\begin{figure*}[htp]
6        \subfloat[]{\includegraphics[scale=0.34]{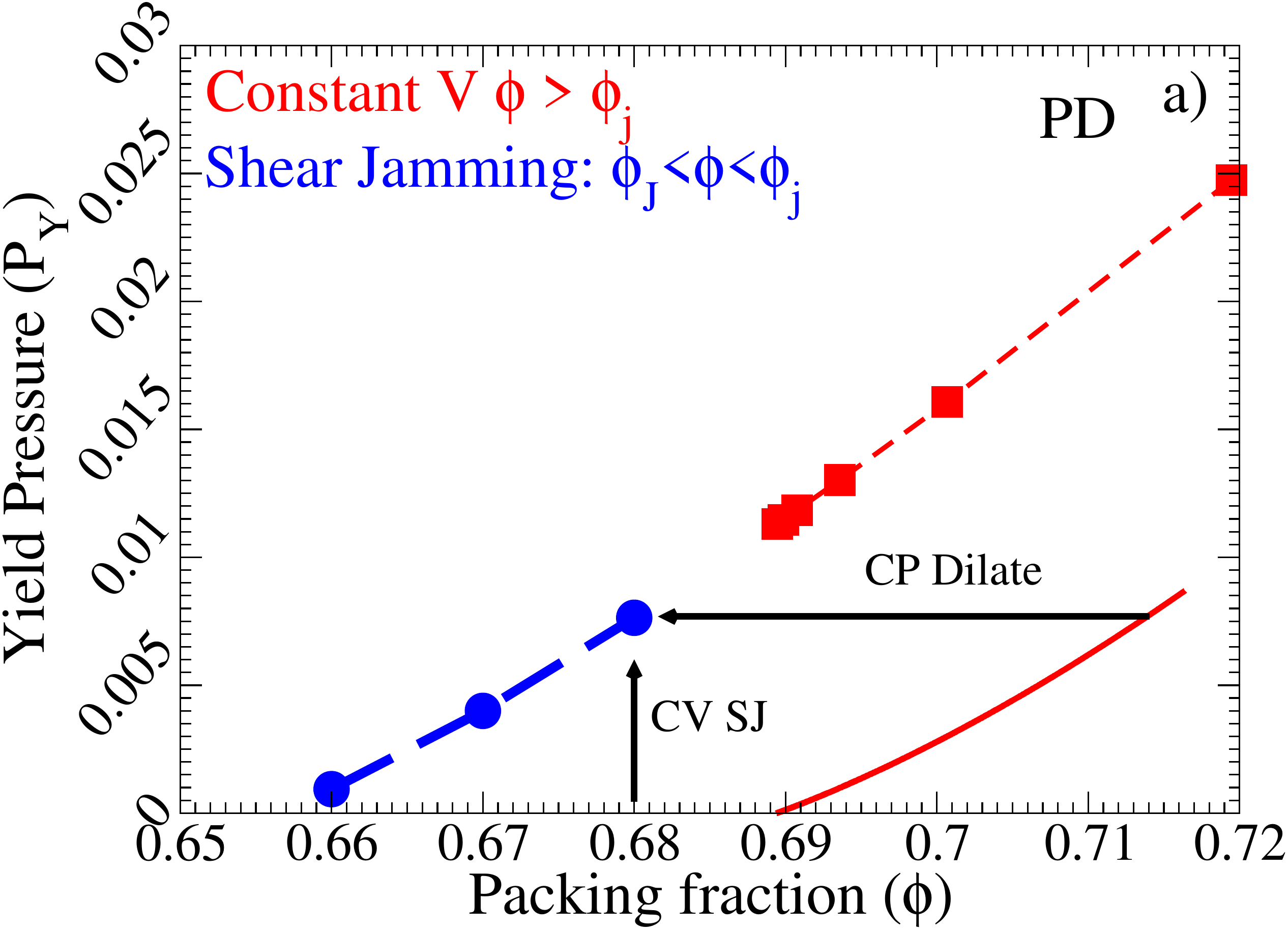}}
        \hfill
        \subfloat[]{\includegraphics[scale=0.34]{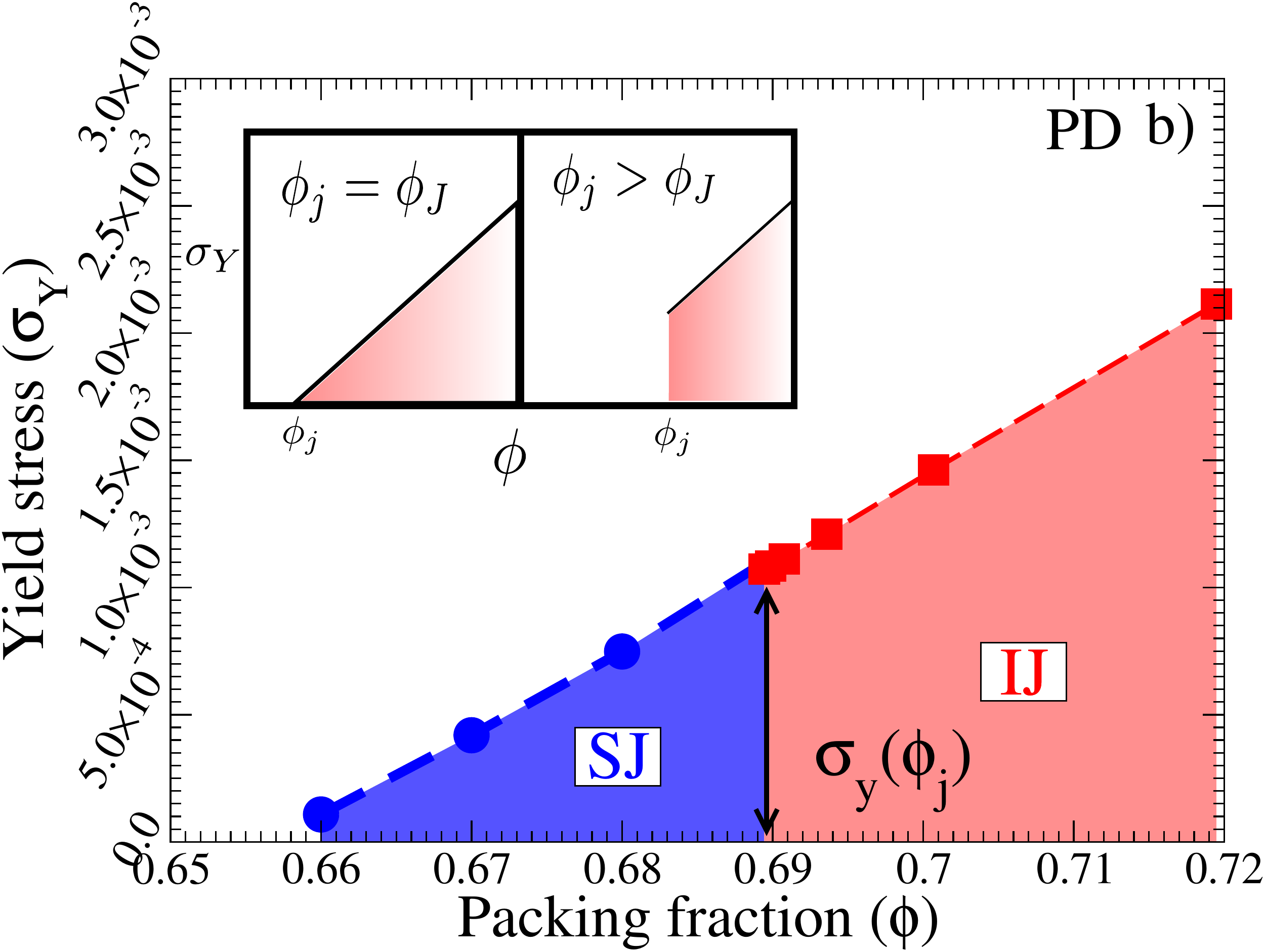}}
        \caption{
        \label{EOS}  
        {
        \bf Generalized zero-temperature jamming phase diagram.
        }
	\textbf{}{a)} Yield pressure $P_Y$ (where $P_Y = P_s$) and \textbf{b)} yield stress $\sigma_Y$ (where $\sigma_Y = \sigma_s$) as functions of volume fraction $\phi$, obtained from constant volume shear of PD systems above $\phi_j=0.689$ (red points) and below (blue points). ({see SI Fig.~\ref{Shear_jamming} for similar plots for the BD model}). 
        Above $\phi_j$, the system is initially jammed by the isotropic compression (IJ), and remains jammed under shear as long as $\sigma_{xz} < \sigma_{Y}$, while below $\phi_j$, the initially unjammed system is shear jammed (SJ) at $\gamma_{j}$ (Fig.~\ref{Details_of_uniform}), and becomes unjammed again (yields)  once $\sigma_{xz}$ reaches $\sigma_{Y}$.
        The isotropic compression EOS $P_{iso}(\phi)$  is also  plotted (red line). The same data are shown in Fig.~\ref{steady_state_cp}(a) and (b) in rescaled plots.
        {
        }
        }
\end{figure*}

\clearpage
\renewcommand{\thetable}{S\arabic{table}}  
\renewcommand{\thefigure}{S\arabic{figure}} 
\setcounter{figure}{0}
\section{ Is friction essential for dilatancy and shear jamming in granular matter? (Supplementary Information) }




\subsection{Additional data for shear jamming}
\label{sec:shear_jamming}


The unjammed configurations at a density $\phi$, where $\phi_J < \phi < \phi_j$, undergo shear jamming when subjected to steady shear at constant volume. 
Shear jamming can be detected by a sharp increase in the stress $\sigma_{xz}$ and in the coordination number $Z_{NR}$ with increasing strain, as shown in Fig. 1.
Additionally, FIG ~\ref{Shear_jamming_p_PE} shows how  the pressure $P$ and the potential energy $PE$ increase with strain. 
In the mechanical annealing protocol, the shear jamming strain $\gamma_j$, which is indicated by an abrupt jump of the pressure $P$ in FIG \ref{fig:Details-of-uniform}, is always
greater than $\gamma_{max} = 0.07$, the training amplitude used in the cyclic shearing.

We also calculate the macroscopic 
friction $\mu=\sigma_{xz}/P$ of the configurations as a function of $\gamma-\gamma_j$ (FIG \ref{macro_friction_rescaled}), which shows a peak in the cases when 
there is a significant overshoot in the stress-strain curve (Fig. 1). This peak, appearing  after the shear jamming strain $\gamma_j$, also exists  in the uniform shear of over-compressed systems ({FIG \ref{macr_fric}}). In both cases, the peak occurs near the yielding point.


\begin{figure*}
    \centering
    \subfloat[]{\includegraphics[scale=0.34]{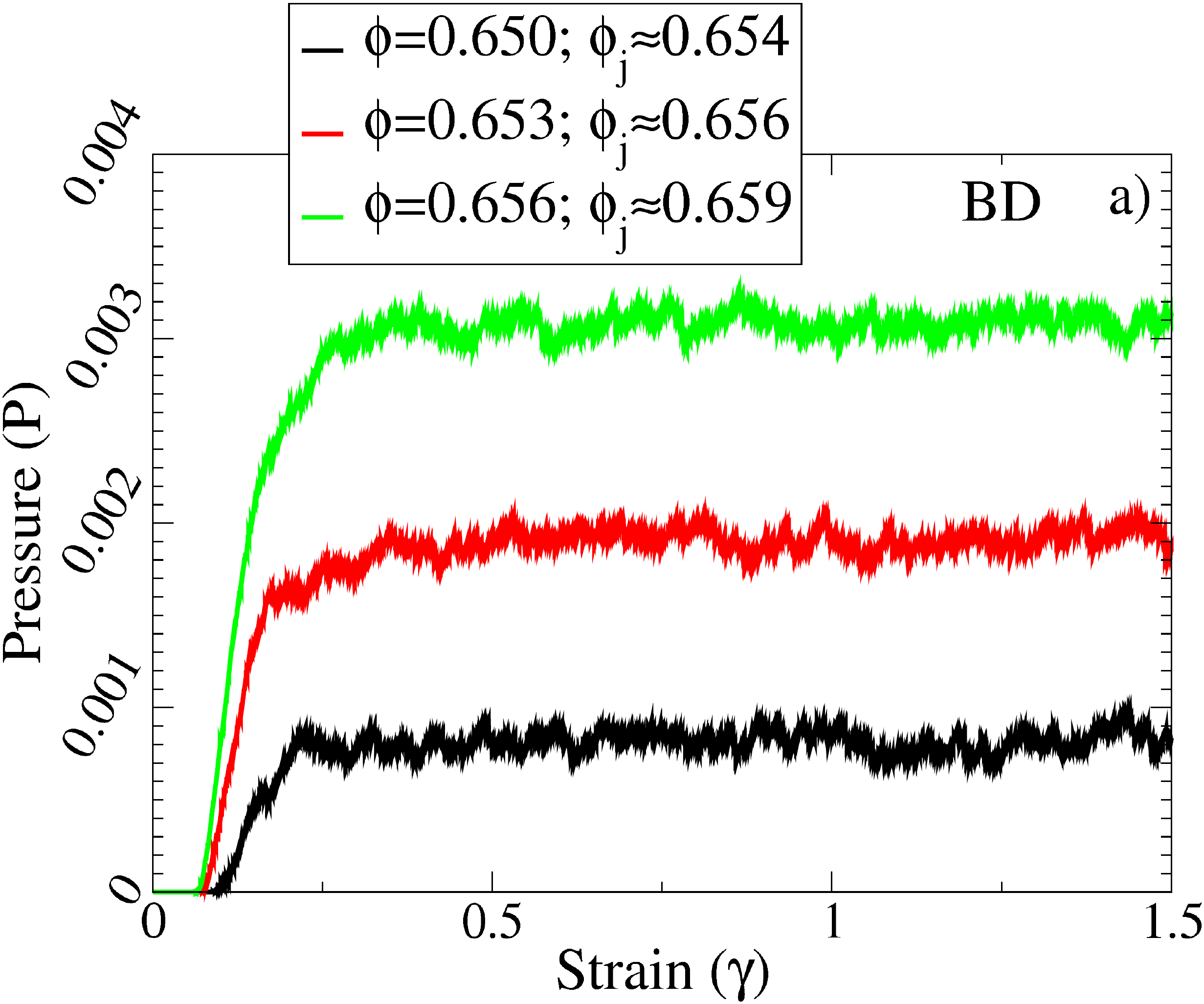}}
    \hfill
    \subfloat[]{\includegraphics[scale=0.34]{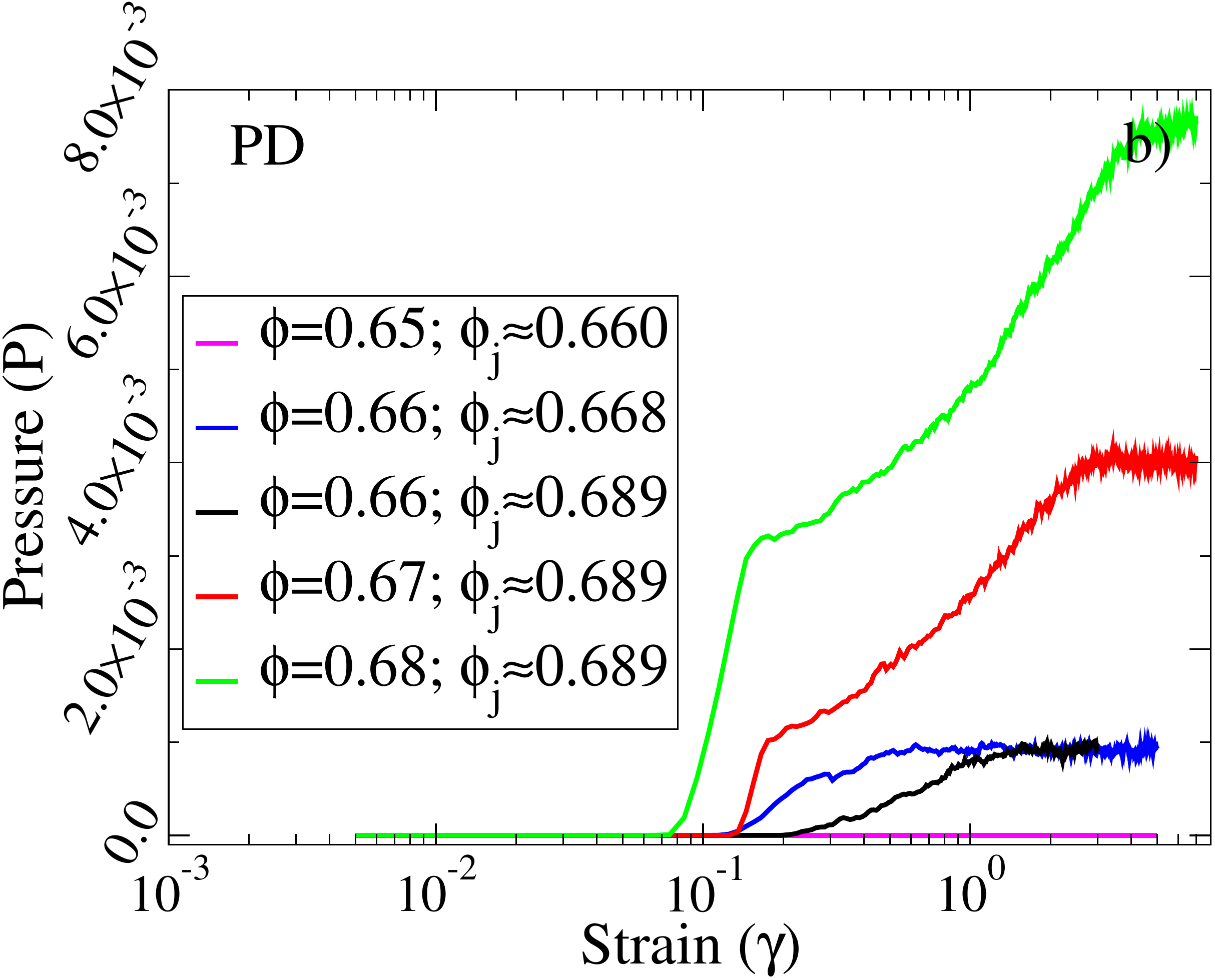}}
    \vfill{}
    \subfloat[]{\includegraphics[scale=0.34]{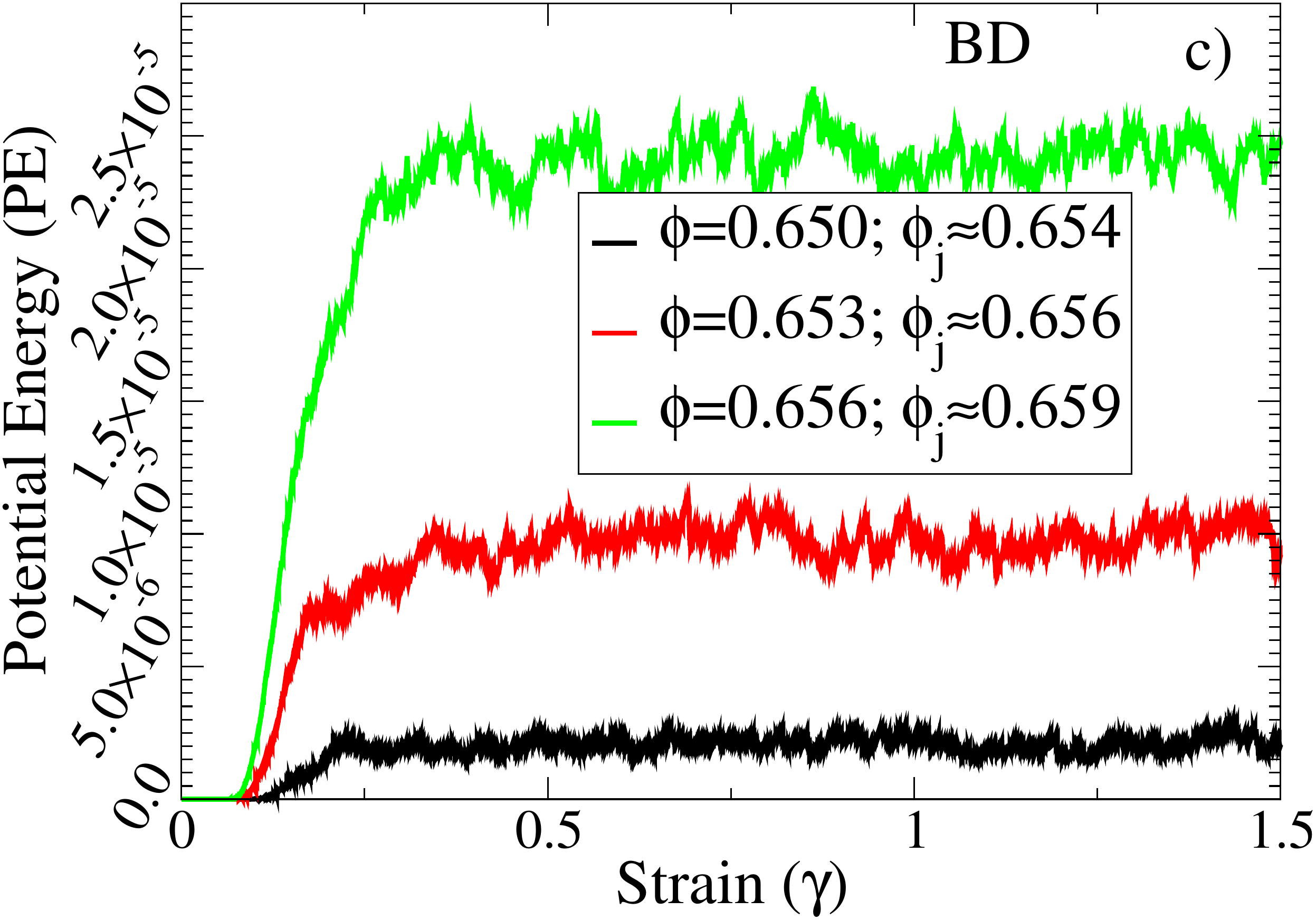}}
    \hfill
    \subfloat[]{\includegraphics[scale=0.34]{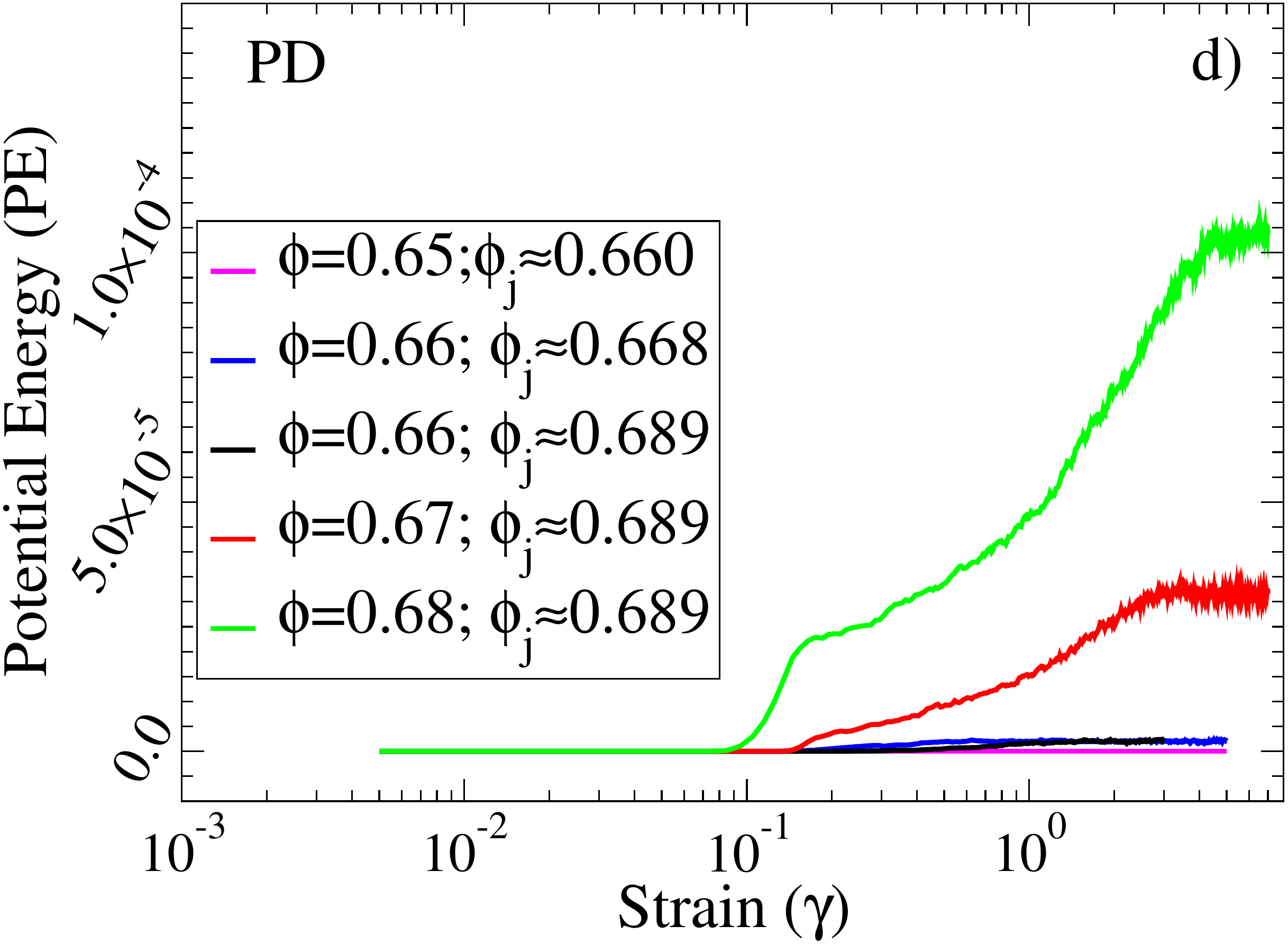}}
    \caption{\textbf{Evolutions of (a-b) pressure $P$ and (c-d) potential energy $PE$ with strain $\gamma$ during shear jamming.}  The constant volume uniform athermal quasi-static shear (AQS) is applied. Data are presented for a few different $\phi$ and $\phi_j$, obtained in both bi-disperse (BD) and poly-disperse (PD) systems. 
    }
    \label{Shear_jamming_p_PE}
\end{figure*}

\begin{figure*}
	{\includegraphics[scale=0.34]{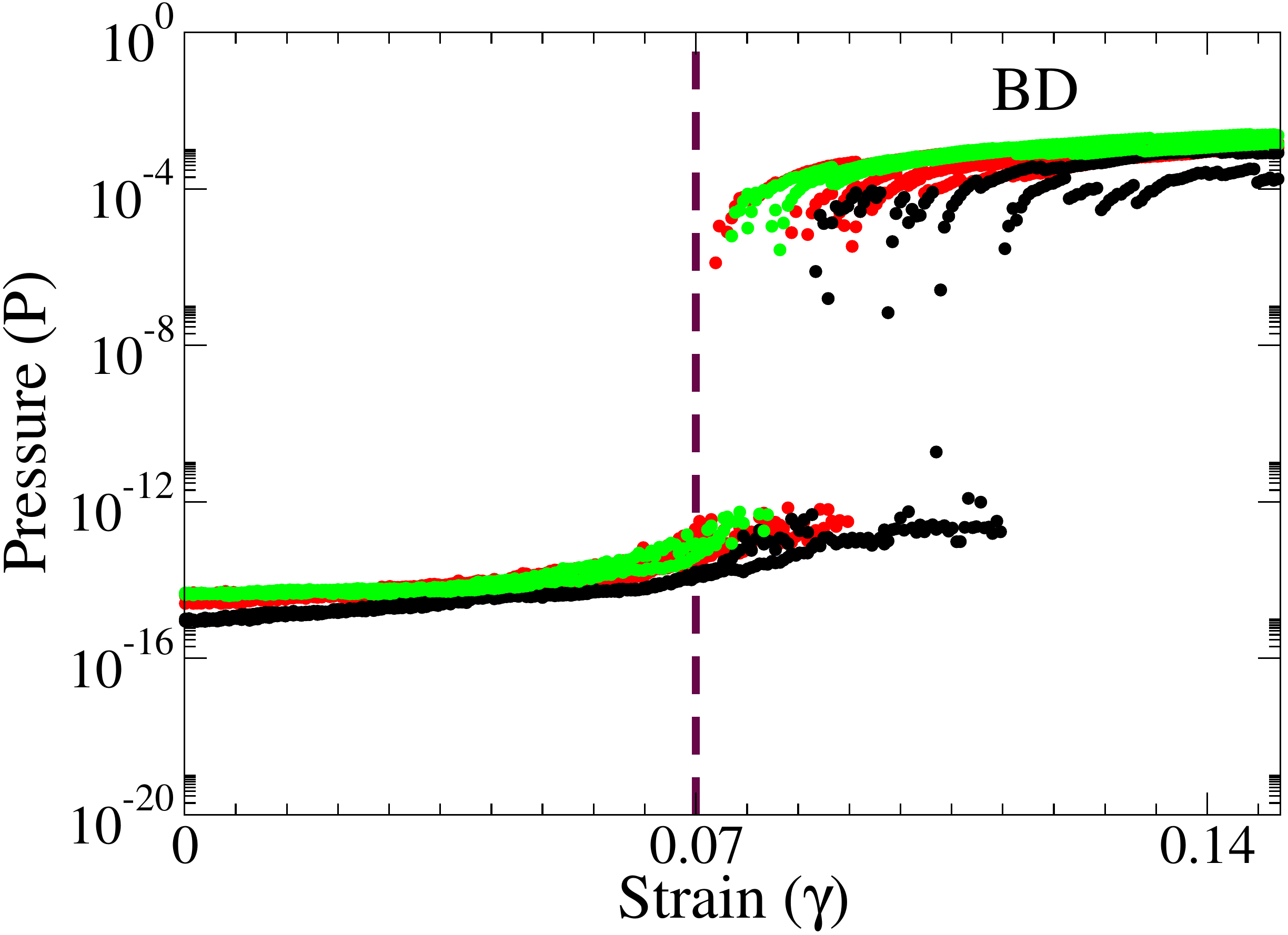}\label{pressure_uniform_ind_samples}}
	\caption{\label{fig:Details-of-uniform}
	{\textbf{Shear jamming under uniform shear in mechanically annealed BD systems.}}
     The pressure increases abruptly as the system is strained beyond $\gamma=\gamma_{max}$, indicating shear jamming. We present multiple realizations for each $\phi_j$, where $\phi_j = 0.659$ (green), 0.656 (red), and 0.654 (black). The densities at which shear is carried out are $\phi=0.656$(green), 0.653 (red), and 0.650 (black). 
     } 
\end{figure*}

\begin{figure*}
	\subfloat[]{\includegraphics[scale=0.34]{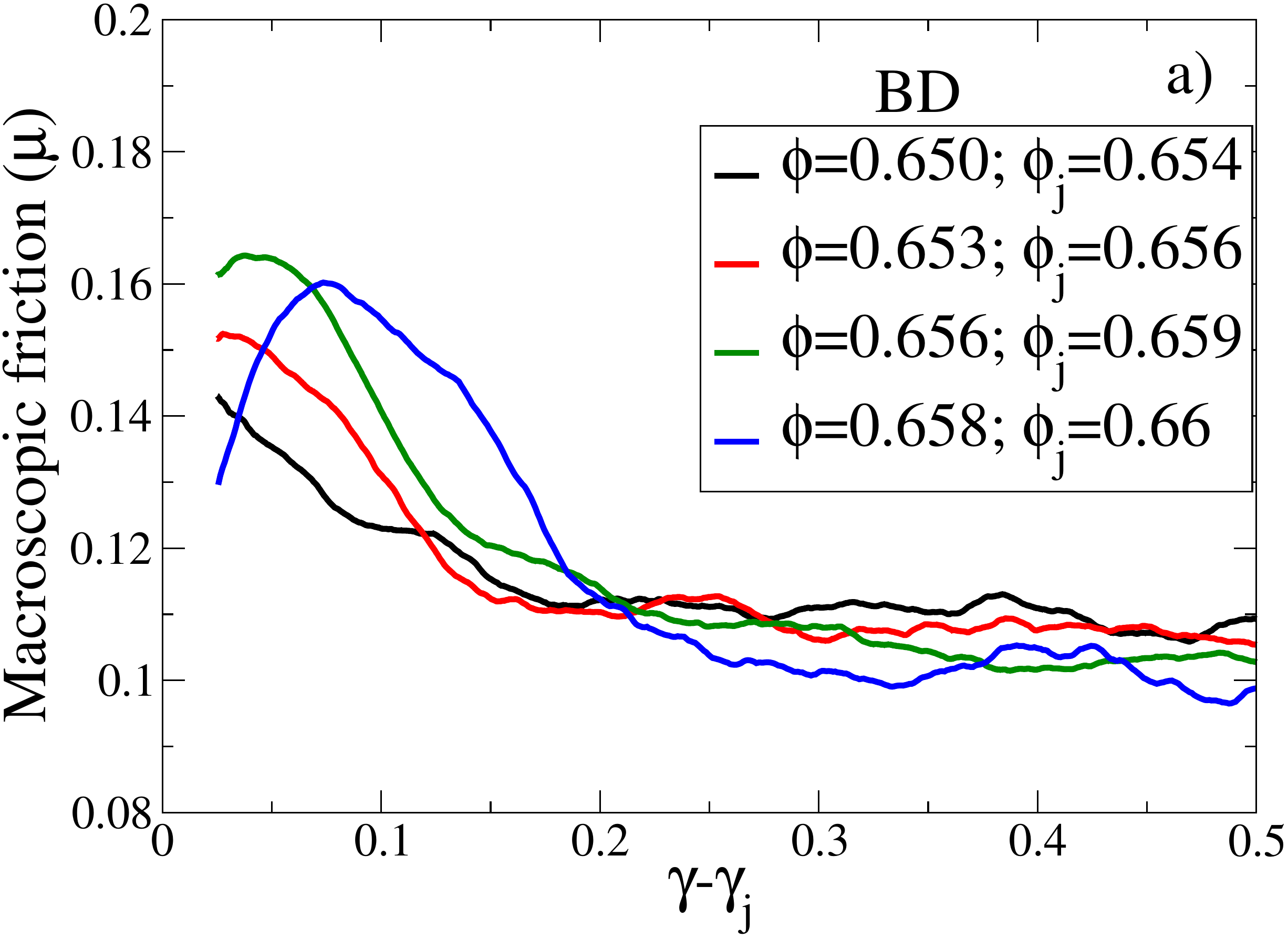}}
	\subfloat[]{\includegraphics[scale=0.34]{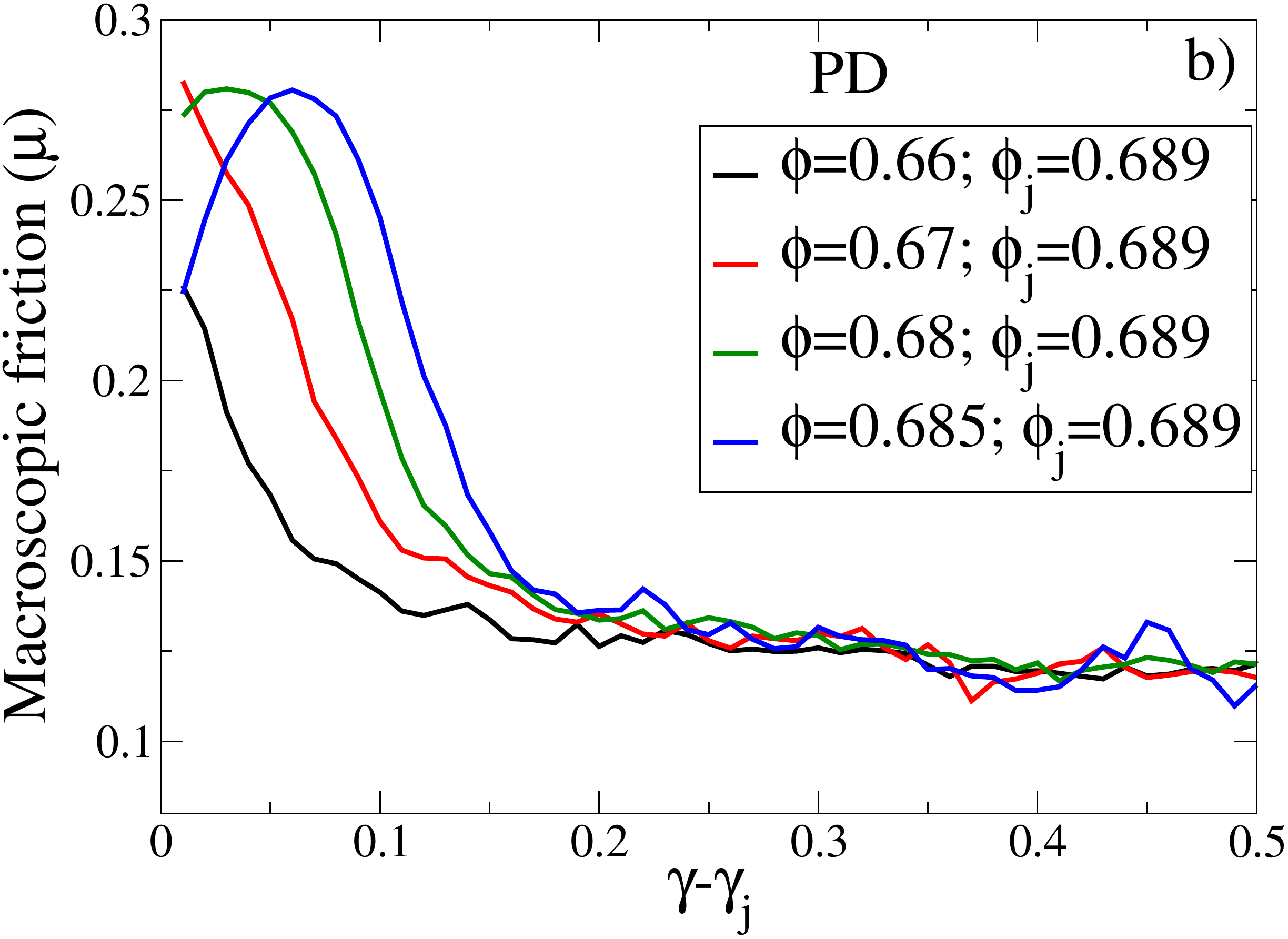}}
		\caption{\label{macro_friction_rescaled}\textbf{Macroscopic friction $\mu$ as a function of distance from jamming strain $\gamma-\gamma_j$ for a) BD and b) PD systems.} The jamming strain $\gamma_j$  is identified as the strain at which $\sigma_{xz}$ increases above $10^{-11}$ in BD systems, and above $10^{-8}$ in PD systems.
	 }
\end{figure*}


\newpage
\subsection{Additional data for the dilatancy effect under constant pressure shear}
Figure~\ref{amount_dilation} shows that, under constant pressure shear deformations,  the amount of dilation $\delta \phi = \phi_{init} - \phi_s$, which is the difference 
between the initial density $\phi_{init}$ and the steady-state density $\phi_s$, increases with the jamming density $\phi_j$ for a fixed pressure $P$, or decreases with $P$ for a fixed $\phi_j$.

\begin{figure*}
	\subfloat[]{\includegraphics[scale=0.34]{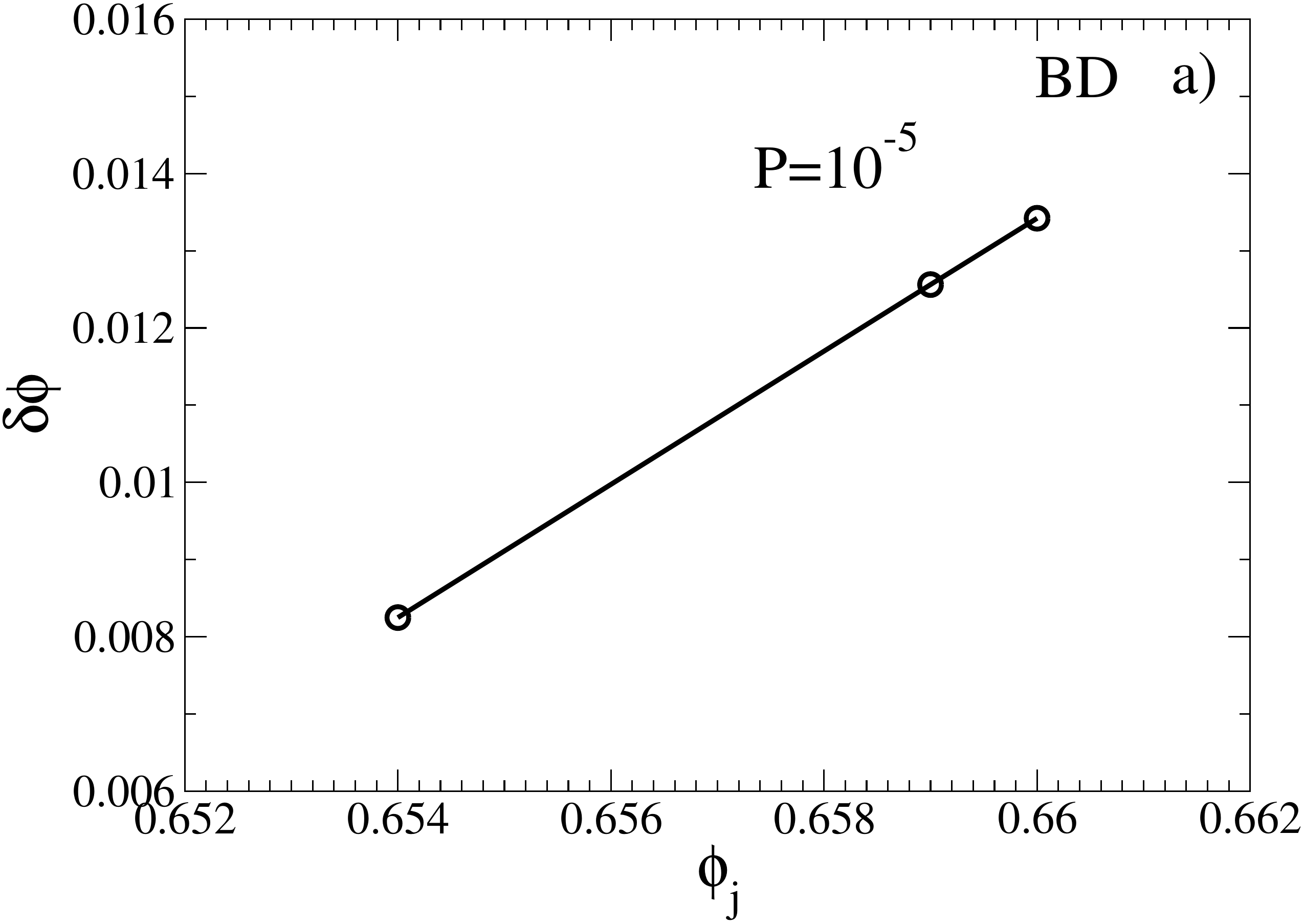}}
	\subfloat[]{\includegraphics[scale=0.34]{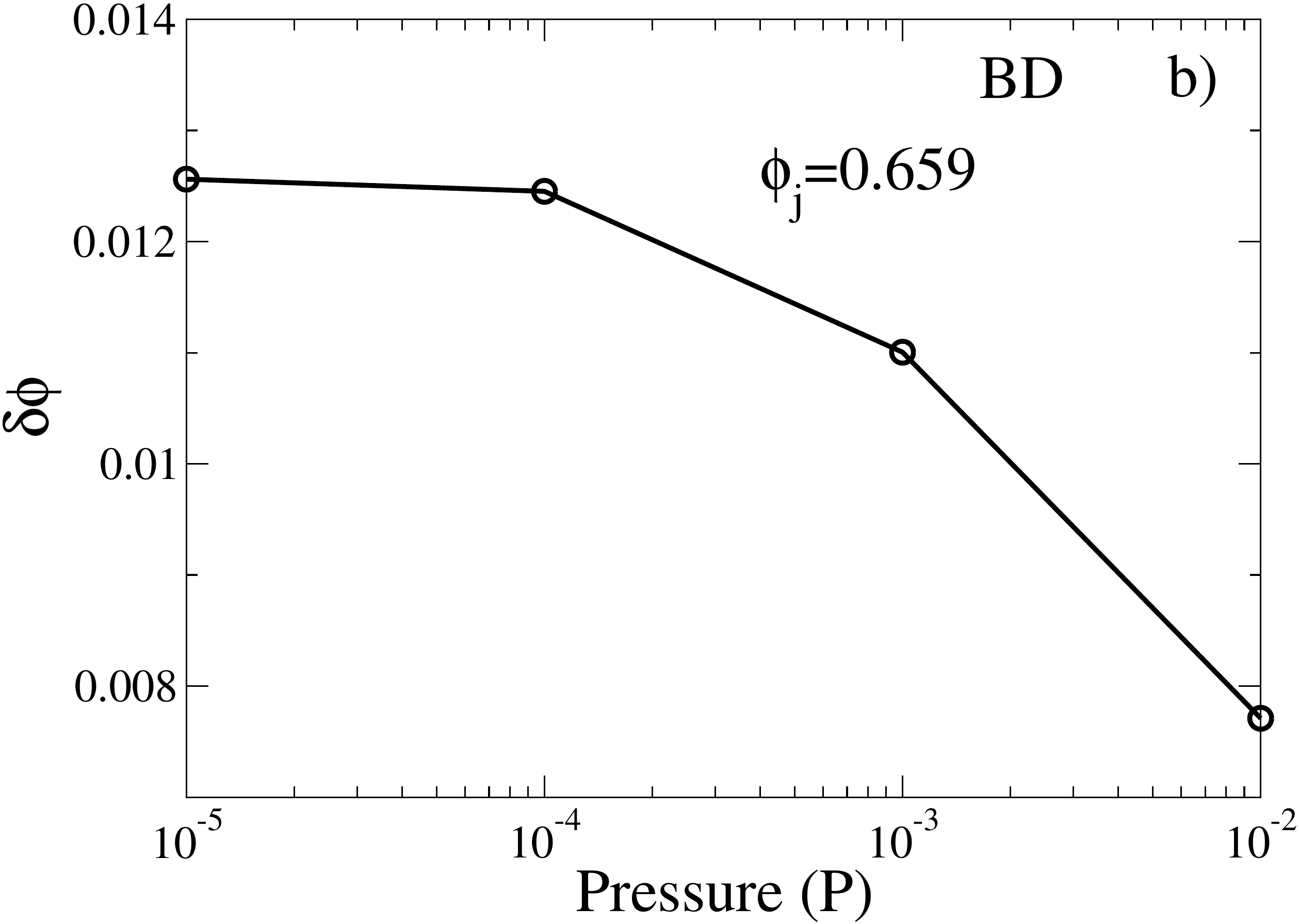}}
	\vfill{}
	\subfloat[]{\includegraphics[scale=0.34]{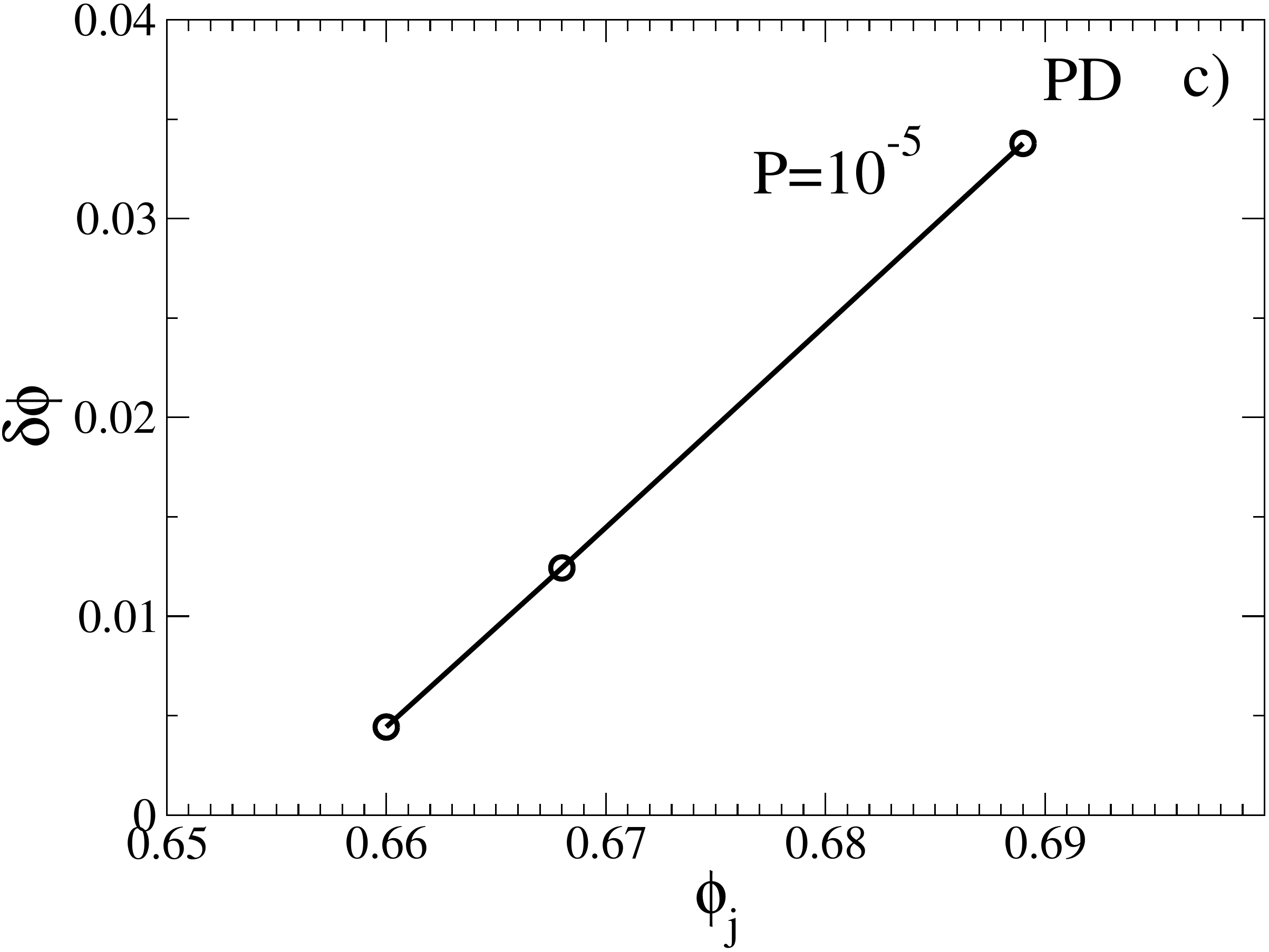}}
	\subfloat[]{\includegraphics[scale=0.34]{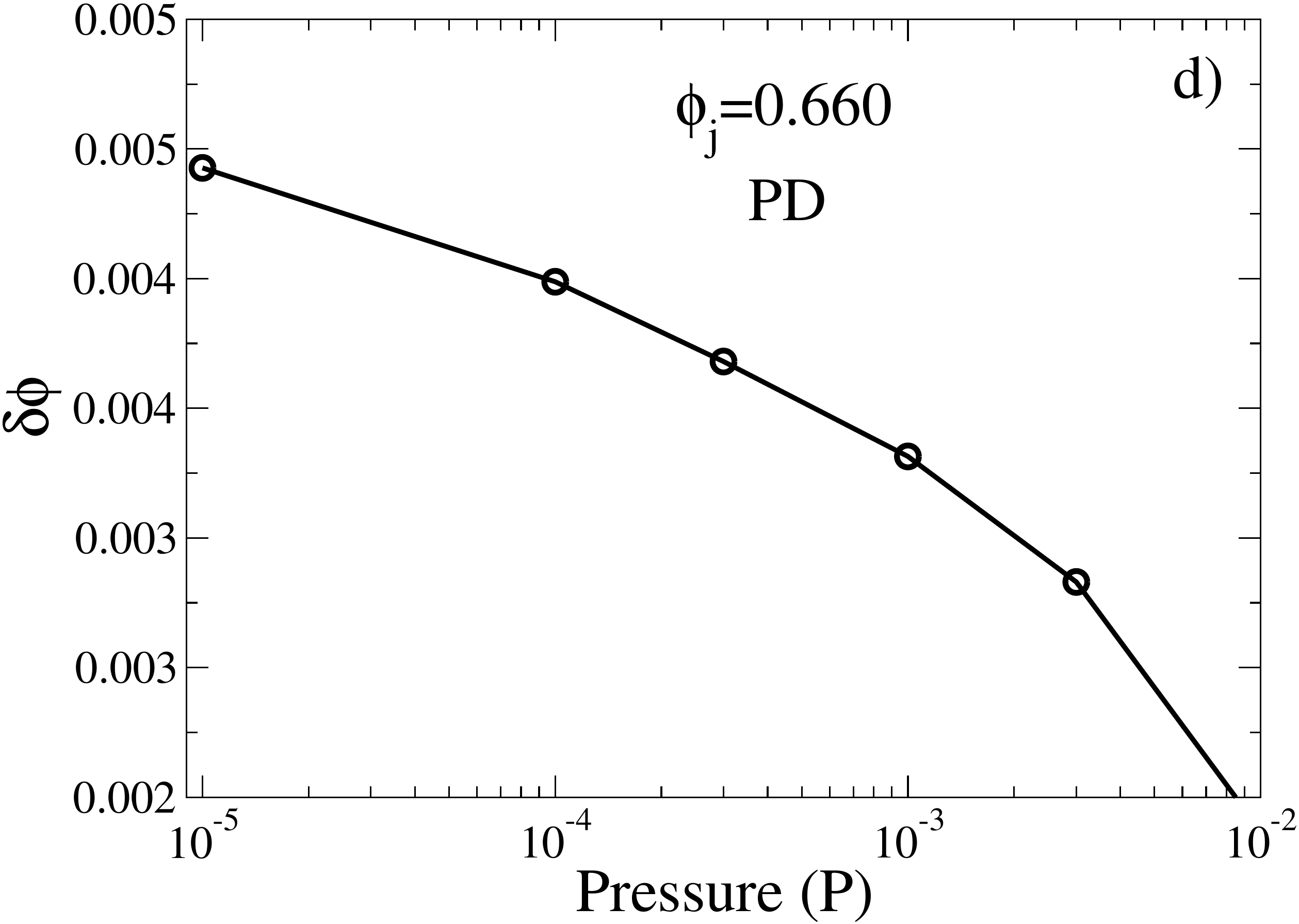}}
	\caption{\label{amount_dilation} \textbf{Amount of dilation in constant pressure shear deformations.} 
	\textbf{a-b)} Amount of dilation $\delta \phi$ as a function of $\phi_j$ for a given pressure $P$,  
	 and \textbf{c-d)}  as a function of $P$ for a given $\phi_j$. 
	 }
\end{figure*}

\subsection{Finite Size analysis of the J-point density $\phi_J$ and the steady-state density $\phi_c$}
We perform a finite size analysis  of both the J-point density $\phi_J$ and the steady-state density $\phi_c$ for the PD model.
Our analysis is based on simulation data obtained from  systems that consist of
$N = 250, 500, 1000, 2000, 4000$ particles, with  256, 192, 128, 64, 64, 64 independent samples respectively.

To estimate the J-point density $\phi_J$, we follow the procedure employed in Ref \cite{coslovich2017exploring}:
starting from a random initial configuration, the system is compressed and decompressed  iteratively,  followed with energy minimization after each step, until reaching the jamming/unjamming threshold where the energy is infinitesimally  positive. More specifically, the particles are inflated instantly to increase the volume fraction by $\delta \phi = 5\times10^{-4}$ during each compression step.
After that, we minimize the energy of the system using the FIRE algorithm~\cite{plimpton1995fast}. If the system is jammed (the residual potential energy per particle after minimization is larger than $10^{-16}$), we decrease $\delta \phi$ by a factor of 2 and decompress the system until it becomes unjammed. We perform a series of decompression and compression as described above, until 
$\delta \phi < 10^{-6}$. Lastly, we perform an additional cycle of compression and decompression:
the compression is performed with $\delta \phi = 10^{-5}$ until the residual energy per particle is larger than $10^{-6}$,  and the decompression is performed with $\delta \phi = 10^{-6}$ until the jammed system becomes unjammed. We identify this unjamming density as $\phi_J$.

To estimate the steady-state density $\phi_c$, we perform constant pressure AQS at  a few different $P_s$, by minimizing the enthalpy using the FIRE algorithm,  and measure the volume fraction $\phi_s(P_s)$ when the stress  reaches a constant value. 
Then we extrapolate $\phi_c$ from  $\phi_s(P_s)$ using the linear relation $\phi_s - \phi_c \sim P_s$ near the zero pressure limit.


The system size dependences of $\phi_J$ and $\phi_c$ are shown in FIG. \ref{finite} a), and the difference $\phi_c - \phi_J$ is plotted as a function of the system size $N$ in FIG. \ref{finite} b). Our results show that $\phi_c$ is always slightly larger than $\phi_J$ in finite size systems, but the difference decreases with $N$. In this paper, we regard $\phi_J \simeq \phi_c$ in the thermodynamical limit $N \to \infty$. 
However, note that several previous studies~\cite{heussinger2010fluctuations, zheng2018jamming, vaagberg2011glassiness} in two dimensions suggested that this difference remains finite (around 0.001-0.002), even in the thermodynamical limit. We do not exclude such a possibility in three dimensions based on our data.

\begin{figure*}
	\subfloat[]{\includegraphics[scale=0.3]{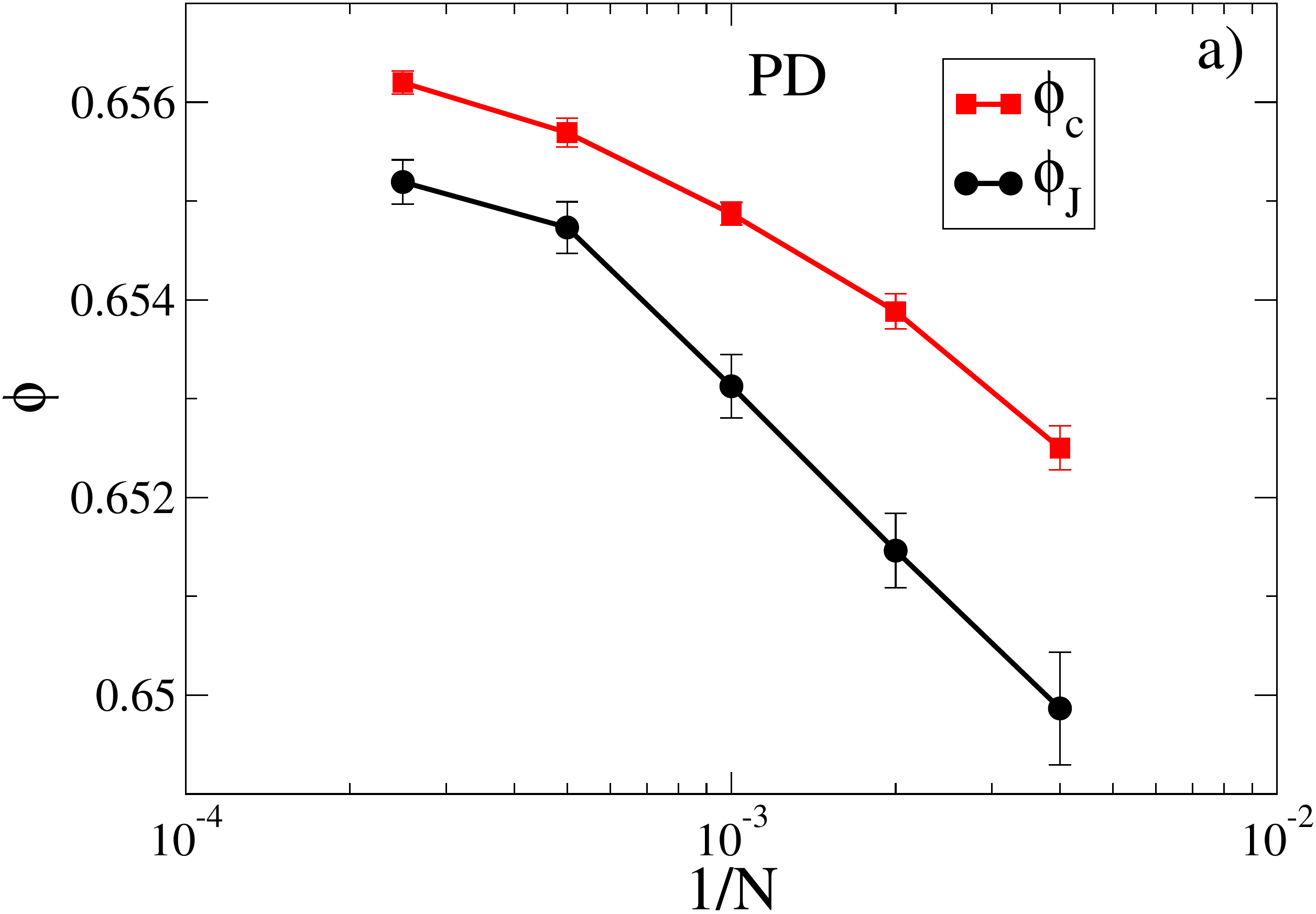}}
	\subfloat[]{\includegraphics[scale=0.3]{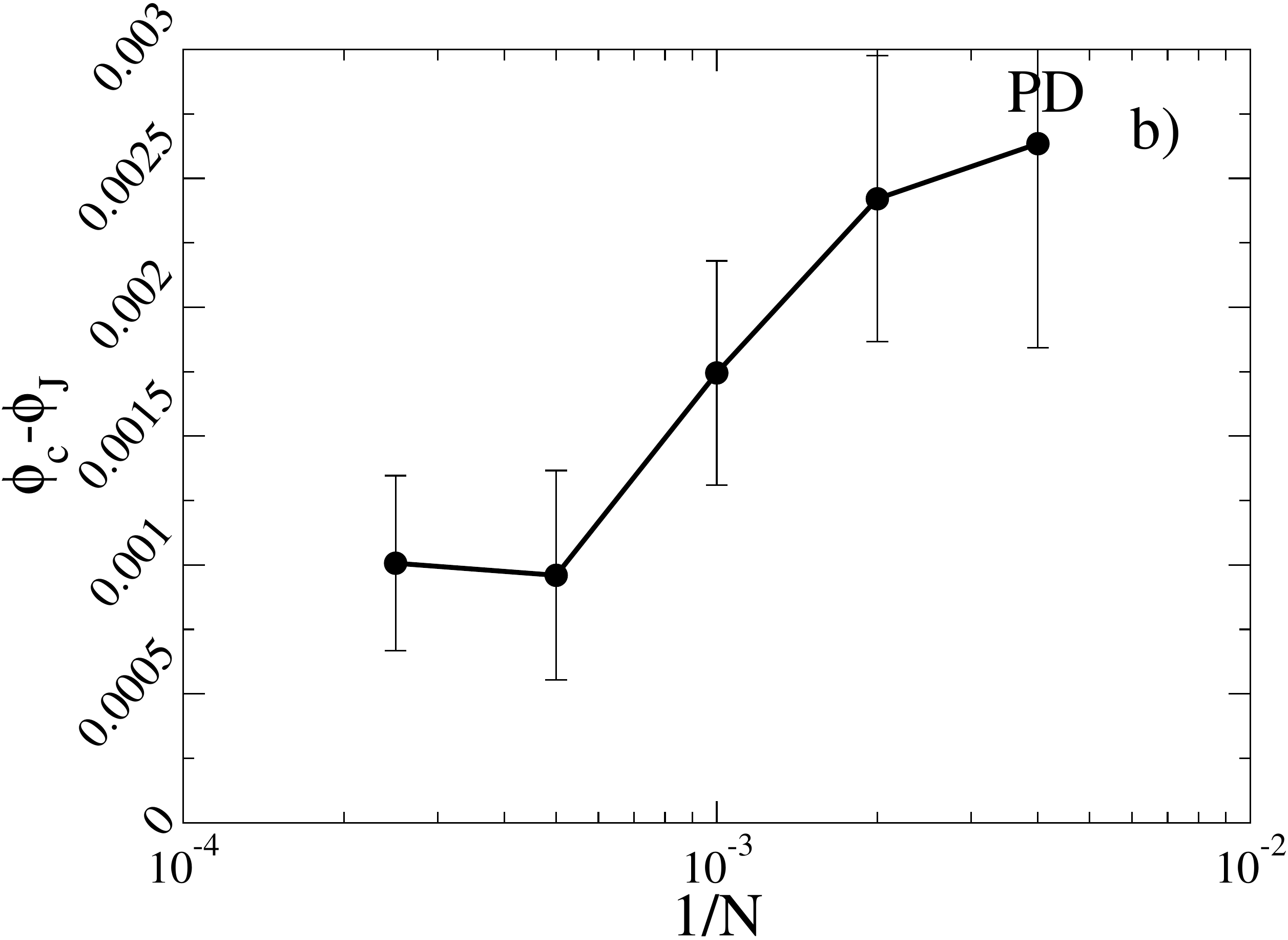}}
	\vfill{}

	\caption{\label{finite}\textbf{Finite size analysis of $\phi_J$ and $\phi_c$ for the PD model.} 
	\textbf{a)} Densities $\phi_J$ and $\phi_c$ as functions of $1/N$. \textbf{b)} The difference $\phi_c - \phi_J$ is plotted as a function of $1/N$.
	The error bars represent  standard errors. }
\end{figure*}


\subsection{Dilatancy effect revealed by pressure increase under constant volume shear}
For over-compressed systems with a jamming density $\phi_j$ above $\phi_J$, the pressure $P$ increases under constant volume shear deformations, which is an effect equivalent to dilatancy in constant pressure shear.
FIG \ref{P_vs_g} shows how the pressure $P$ increases from $P_{init}$ when the constant volume shear is applied, and 
FIG \ref{macr_fric} shows the evolutions of the macroscopic friction $\mu$.
We find that the peak in macroscopic friction is more prominent for configurations with a larger $\phi_j$. 
The scaling relationship between the steady-state macroscopic friction $\mu_s$ and pressure $P_s$, 
$\mu_s = \mu_0 - c P_s^\beta$, is shown in 
FIG \ref{Steady_macro}. 
Note that, in Fig. 1 of the main text and Sec.~\ref{sec:shear_jamming}, the initial configurations are unjammed ($P_{init} = 0$ or $\phi<\phi_j$). In that case, the constant volume shear deformation firstly jams the system, and then increases the pressure (see FIG~\ref{Shear_jamming_p_PE}).


\begin{figure*}
	\subfloat[]{\includegraphics[scale=0.34]{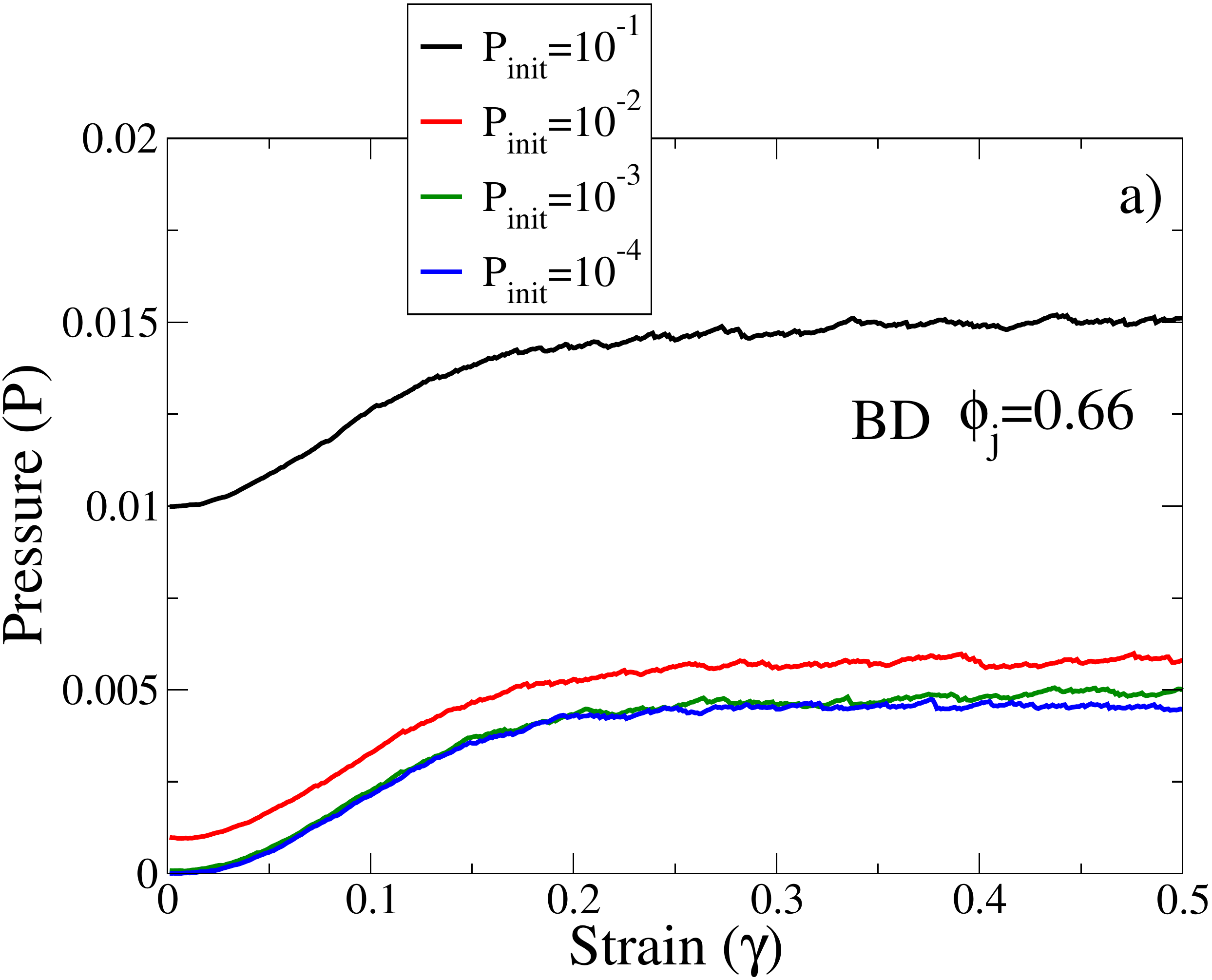}}
	\subfloat[]{\includegraphics[scale=0.34]{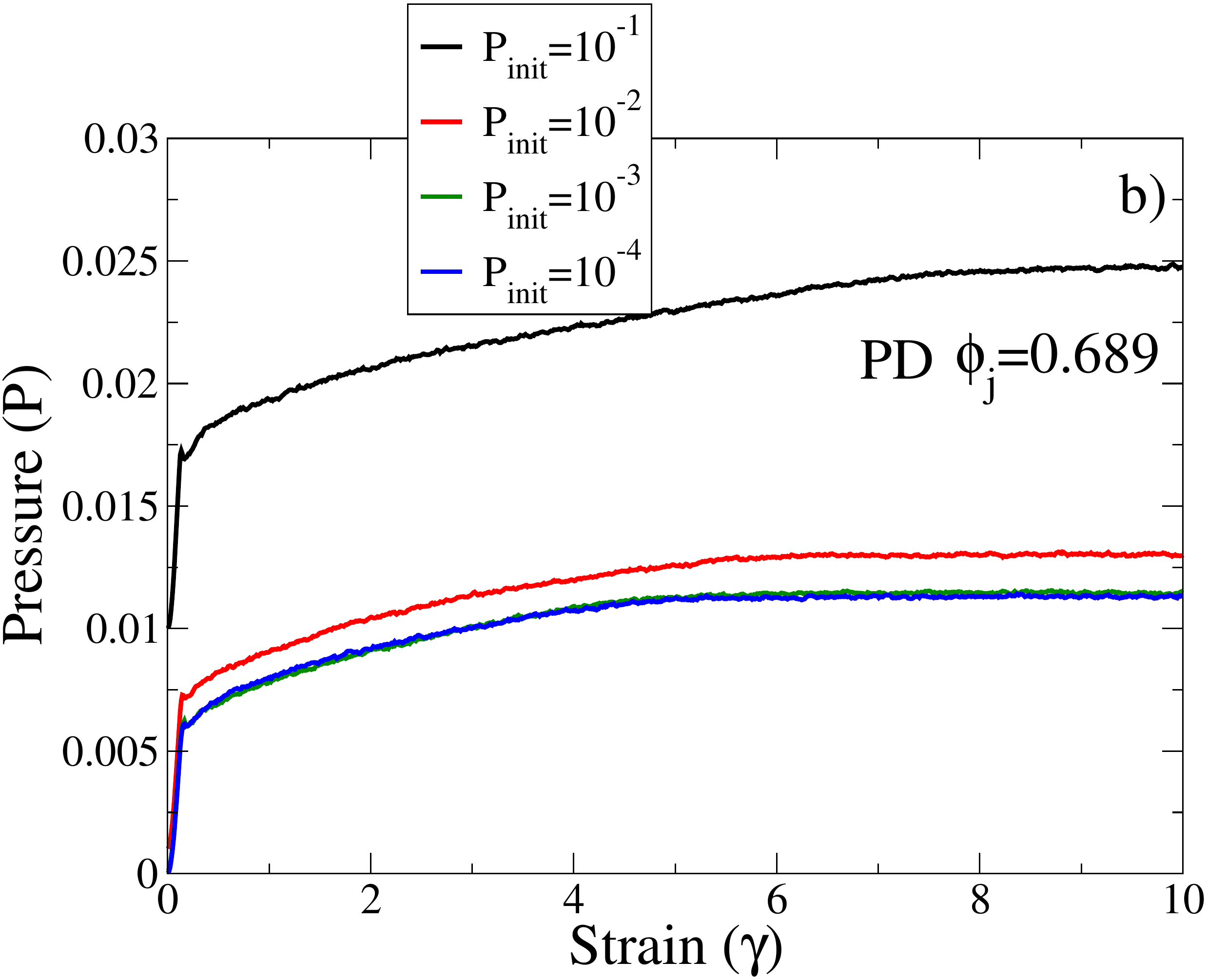}}
		\caption{\label{P_vs_g}\textbf{Pressure $P$ as a function of strain $\gamma$  under constant volume shear deformations, for over-compressed systems. }
           The pressure increases from the initial value $P_{init}$ as the system is strained in both \textbf{a)} BD and \textbf{b)} PD systems.}
\end{figure*}
\begin{figure*}
	{\includegraphics[scale=0.34]{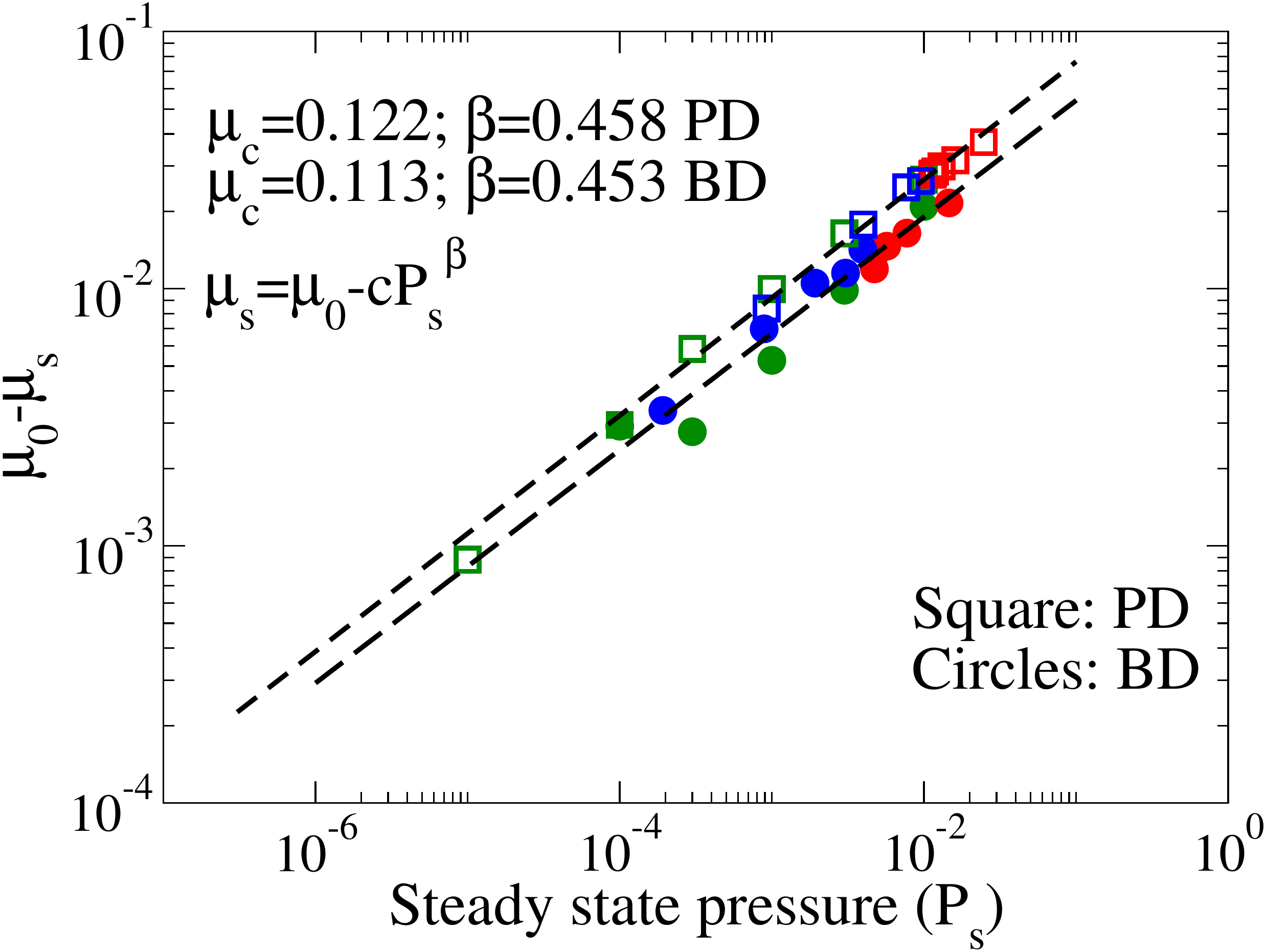}}
	\caption{ \label{Steady_macro}
	 \textbf{Scaling relationship between the steady-state macroscopic friction $\mu_s$ and  pressure $P_s$.} The data for both BD ({$\phi_j \approx 0.660$}) and PD ({$\phi_j \approx 0.689$}) systems are fitted to the empirical scaling form $\mu_s = \mu_0 - c P_s^\beta$ (dashed lines).
	The data obtained from constant pressure shear (green), constant volume shear for $\phi_J< \phi < \phi_j$ (shear jamming, blue), and constant volume shear for $\phi>\phi_j$ (red) are presented.
	} 
\end{figure*}

\begin{figure*}
	\subfloat[]{\includegraphics[scale=0.34]{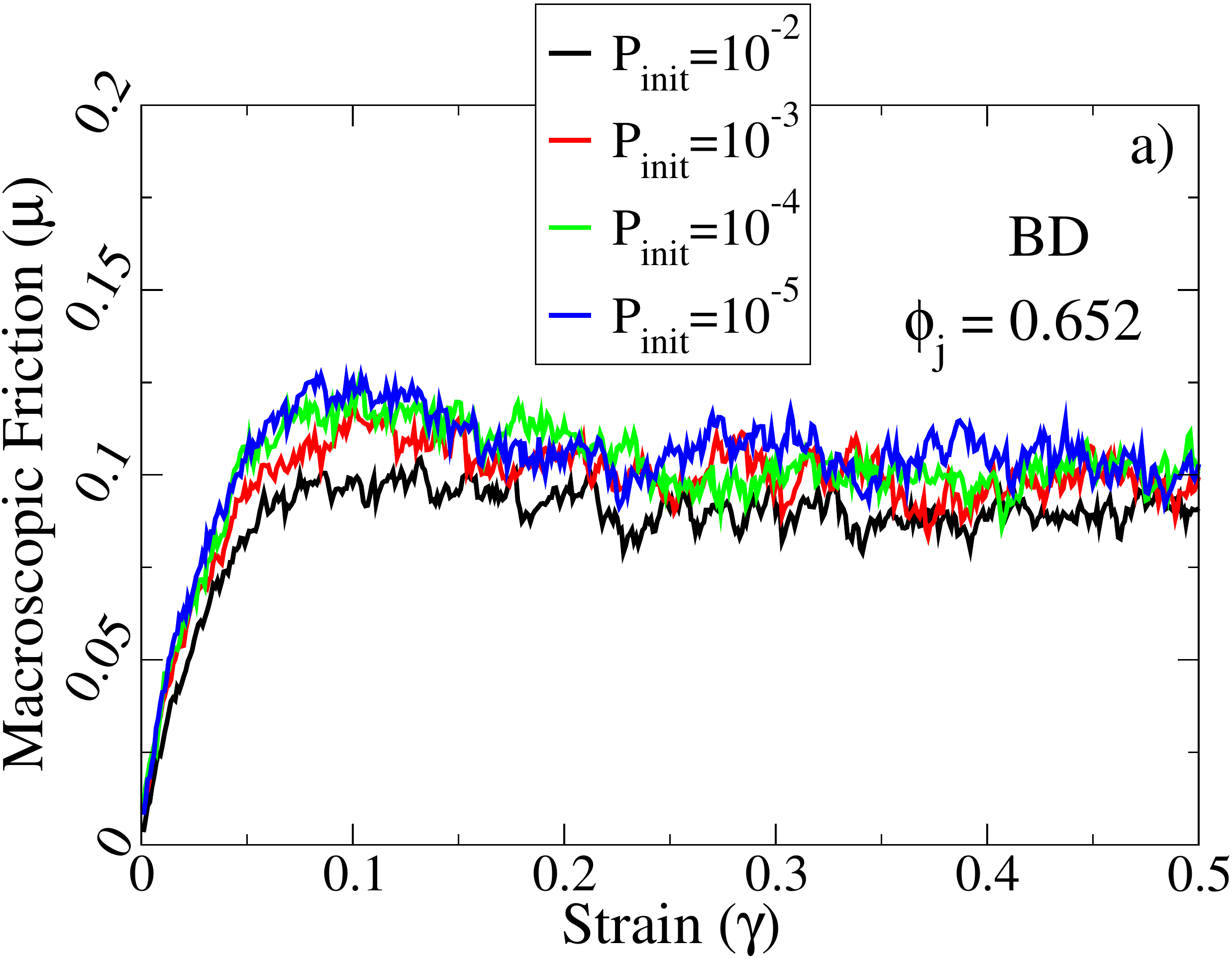}}
	\hfill
	\subfloat[]{\includegraphics[scale=0.34]{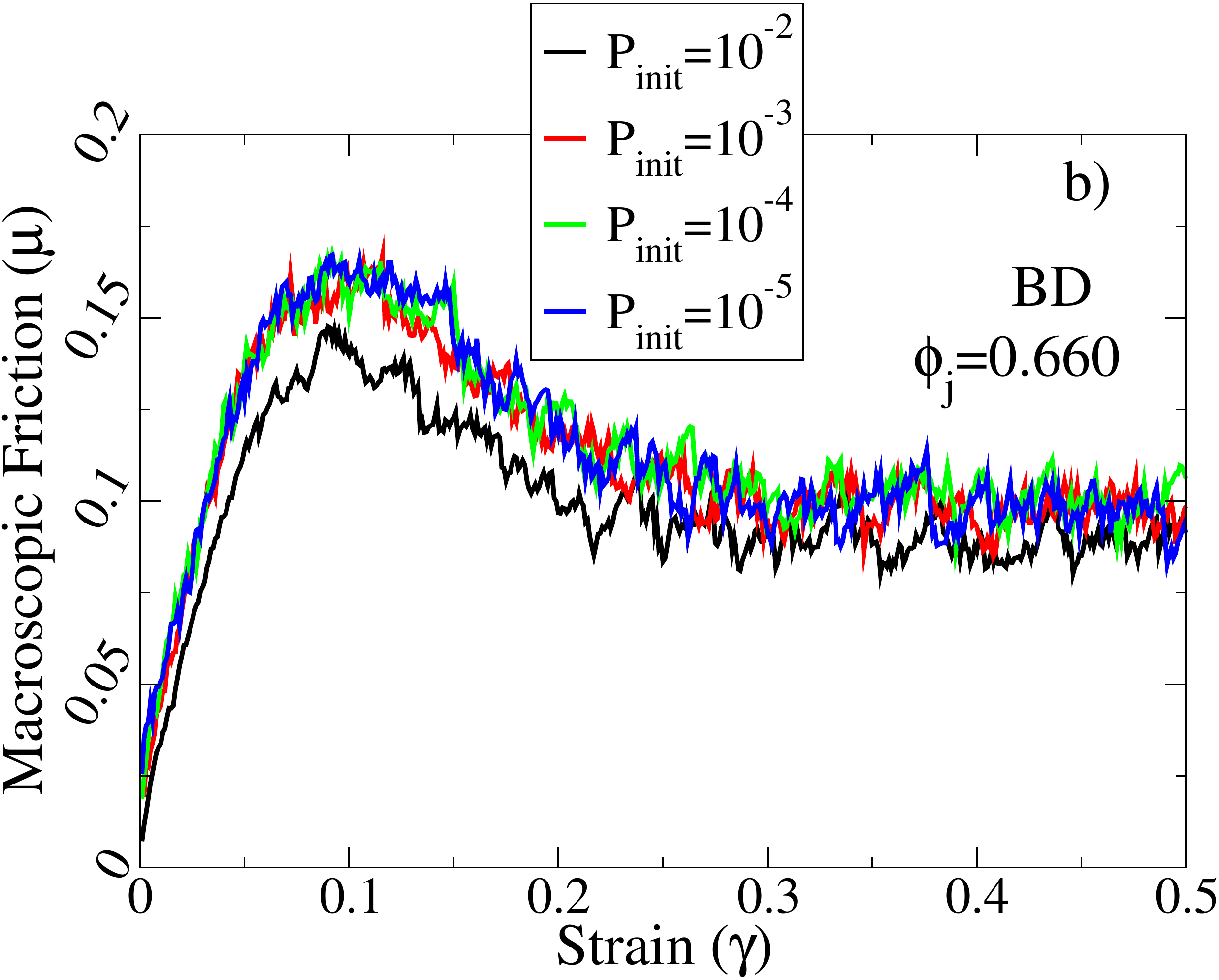}}
	\vfill{}
	\subfloat[]{\includegraphics[scale=0.34]{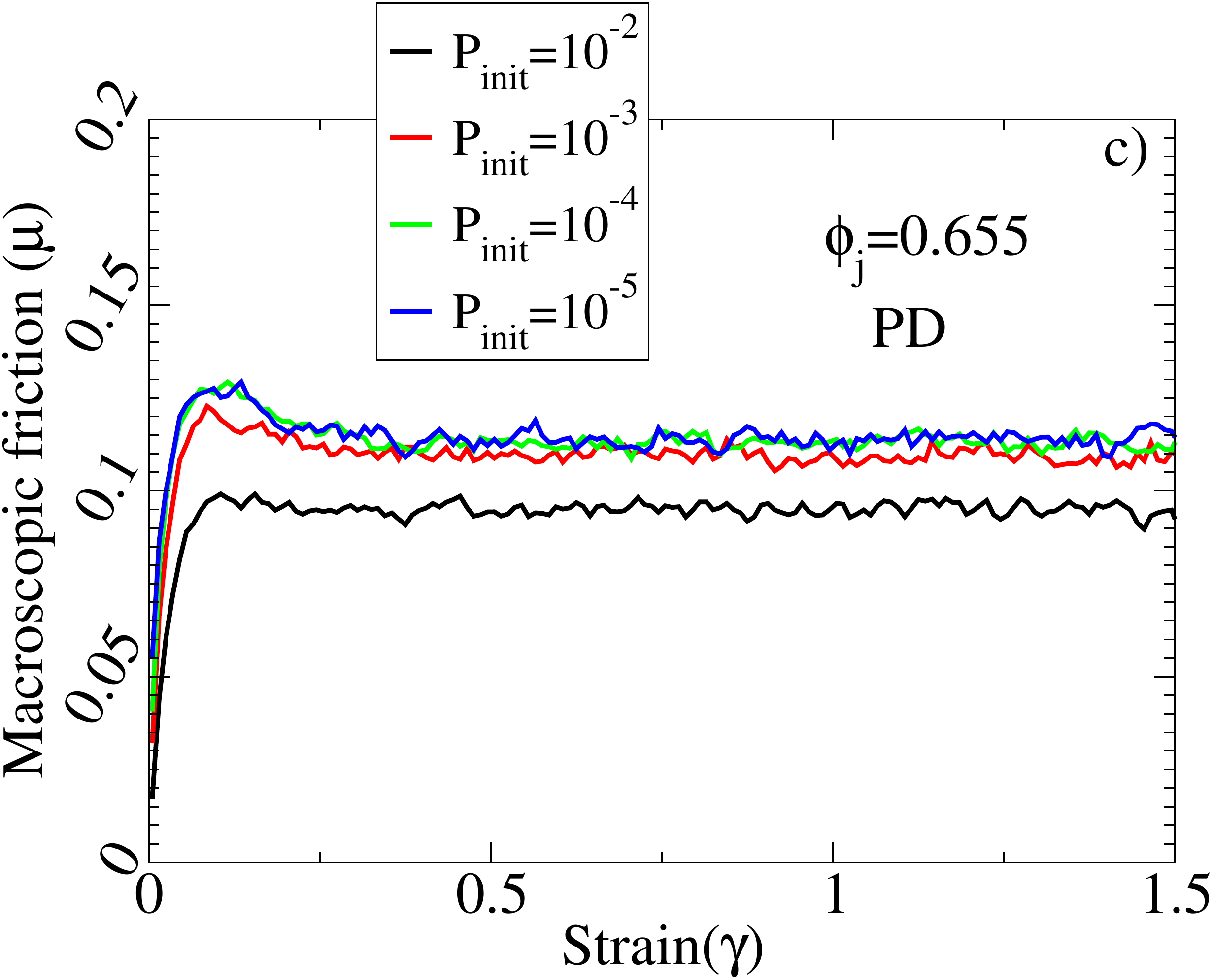}}
	\hfill
	\subfloat[]{\includegraphics[scale=0.34]{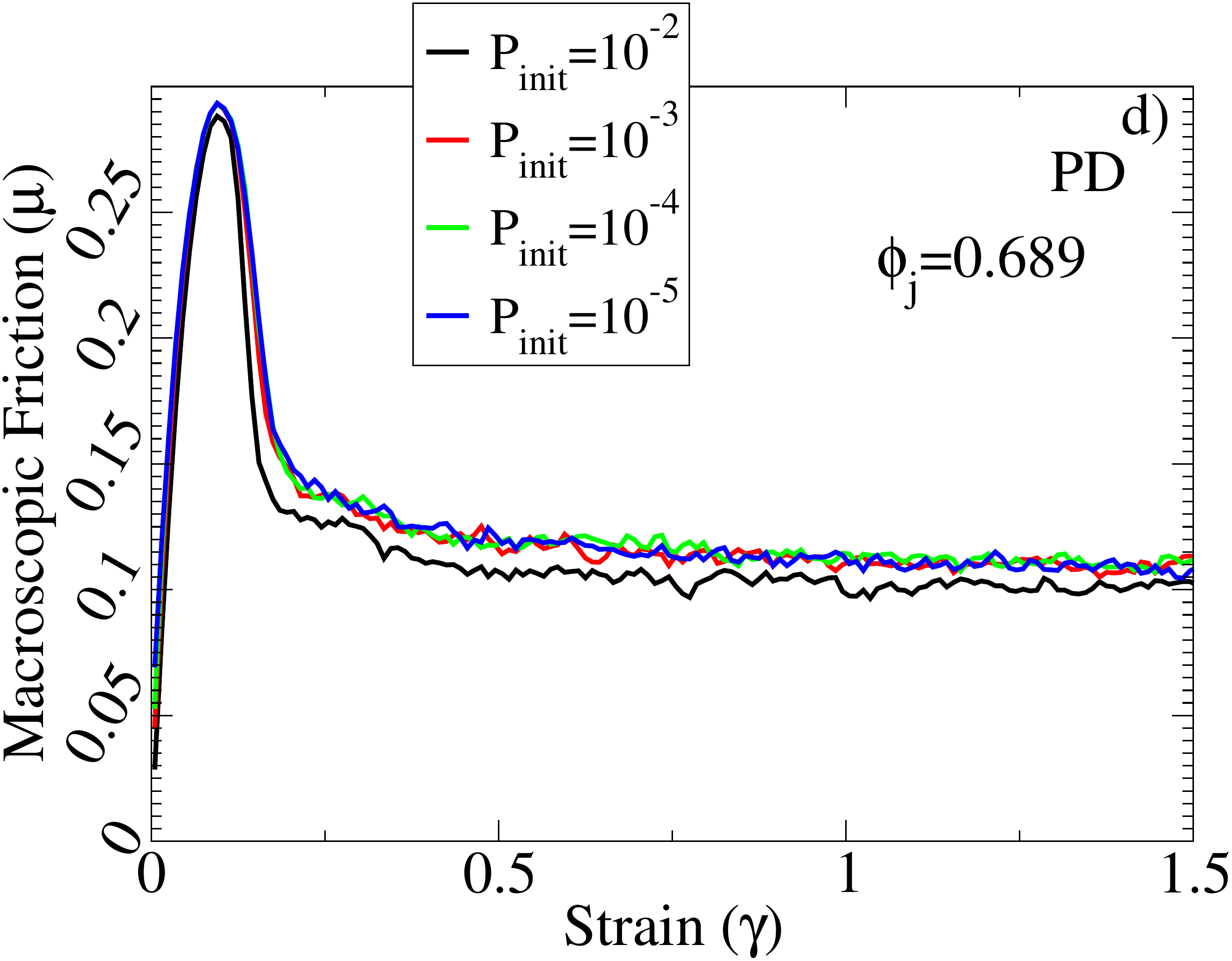}}
	\caption{ 
	    \label{macr_fric}
		\textbf{Microscopic friction $\mu$ of over-compressed systems ($P_{init} > 0$, $\phi > \phi_j$) as a function of strain $\gamma$ under 
		constant volume shear.}
		Data are plotted for two different  $\phi_j$ and four different $P_{init}$ in both BD and PD models. 
		} 
\end{figure*}

%
%


\subsection{Additional data for equations of state}
\subsubsection{Equations of state of steady-states}
Here we explain how to obtain the steady-state equations of state (EOSs) of pressure $P_s(\phi_s)$ and of stress $\sigma_{xz,s}(\phi_s)$. 
For the EOS of pressure,  we firstly calculate the average pressure-strain curve $P(\gamma)=\langle P^{ind}(\gamma) \rangle$ in constant volume shear simulations where the density $\phi = \phi_s$ is fixed, or the average density-strain curve $\phi(\gamma)=\langle \phi^{ind}(\gamma) \rangle$ in constant pressure shear simulations where the pressure $P = P_s$ is fixed. Here 
$P^{ind}(\gamma)$ and $\phi^{ind}(\gamma)$ are the  pressure and density of individual samples at strain $\gamma$, and $\langle \ldots \rangle$ represents the sample average.
We then extrapolate the large-$\gamma$ limits of  $P(\gamma)$ and $\phi(\gamma)$ as the steady-state values $P_s$ and $\phi_s$.
By varying the control parameter $\phi_s$ in constant volume shear, and $P_s$ in  constant pressure shear, we obtain the pressure EOS $P_s(\phi_s)$ for both protocols (FIG \ref{scaling}). 
 The same procedure is applied to get the stress EOS  $\sigma_{xz,s}(\phi_s)$.

To estimate the density $\phi_c$ of the critical state, we fit the EOS data $P_s(\phi_s)$ and $\sigma_{xz,s}(\phi_s)$ 
to the asymptotic linear scalings near the zero pressure limit,
\begin{equation}
    P_s (\phi_s) = P_0 (\phi_s/\phi_c^{P} -1),
    \label{eq:Ps}
\end{equation}
and
\begin{equation}
    \sigma_{xz, s} (\phi_s) = \sigma_0 (\phi_s/\phi_c^{\sigma} -1),
    \label{eq:sigmas}
\end{equation}
where $P_0, \sigma_0, \phi_c^{P}, \phi_c^{\sigma}$ are fitting parameters (see FIG~\ref{scaling}). The values of the fitting parameters are summarized in TABLE~\ref{fit_values}, which show that consistently
$\phi_c^P = \phi_c^\sigma$ within the numerical uncertainty. We therefore determine the critical-state density as $\phi_c = \phi_c^P = \phi_c^\sigma$.


\subsubsection{Equation of state of isotropic-jamming}
We first measure the pressure $P^{ind}_{iso}(\Delta \phi)$ at a given $\Delta\phi = \phi - \phi_J^{ind}$ for each individual sample, where $\phi^{ind}_J$ is the individual sample jamming density determined according to the jamming criterion described in METHODS. To do that, we compress  the configuration from $\phi^{ind}_J$ in small increments of density $\delta \phi=10^{-4}$, up to the target density $\phi > \phi_J^{ind}$. We then average over samples to obtain the EOS, $P_{iso}(\Delta \phi ) = \langle  P^{ind}_{iso}(\Delta \phi)  \rangle $.
The isotropic jamming density $\phi_J$ is determined from the average value of $\phi_J^{ind}$, $\phi_J = \langle \phi^{ind}_J \rangle$. The isotropic jamming EOS satisfies the linear scaling near $\phi_J$,
\begin{equation}
    P_{iso}(\phi) = P_0'(\phi/\phi_J -1),
    \label{eq:Piso}
\end{equation}
where $P'_0 = 0.29$ (BD model) and 0.21 (PD model) are used to 
re-scale $P_{iso}$ such that the isotropic jamming and the steady-sate EOSs collapse onto the universal curve (Fig. 3a). The values of $\phi_J$ and $P'$ are listed in TABLE~\ref{fit_values}.


\begin{figure*}
	\subfloat[]{\includegraphics[scale=0.34]{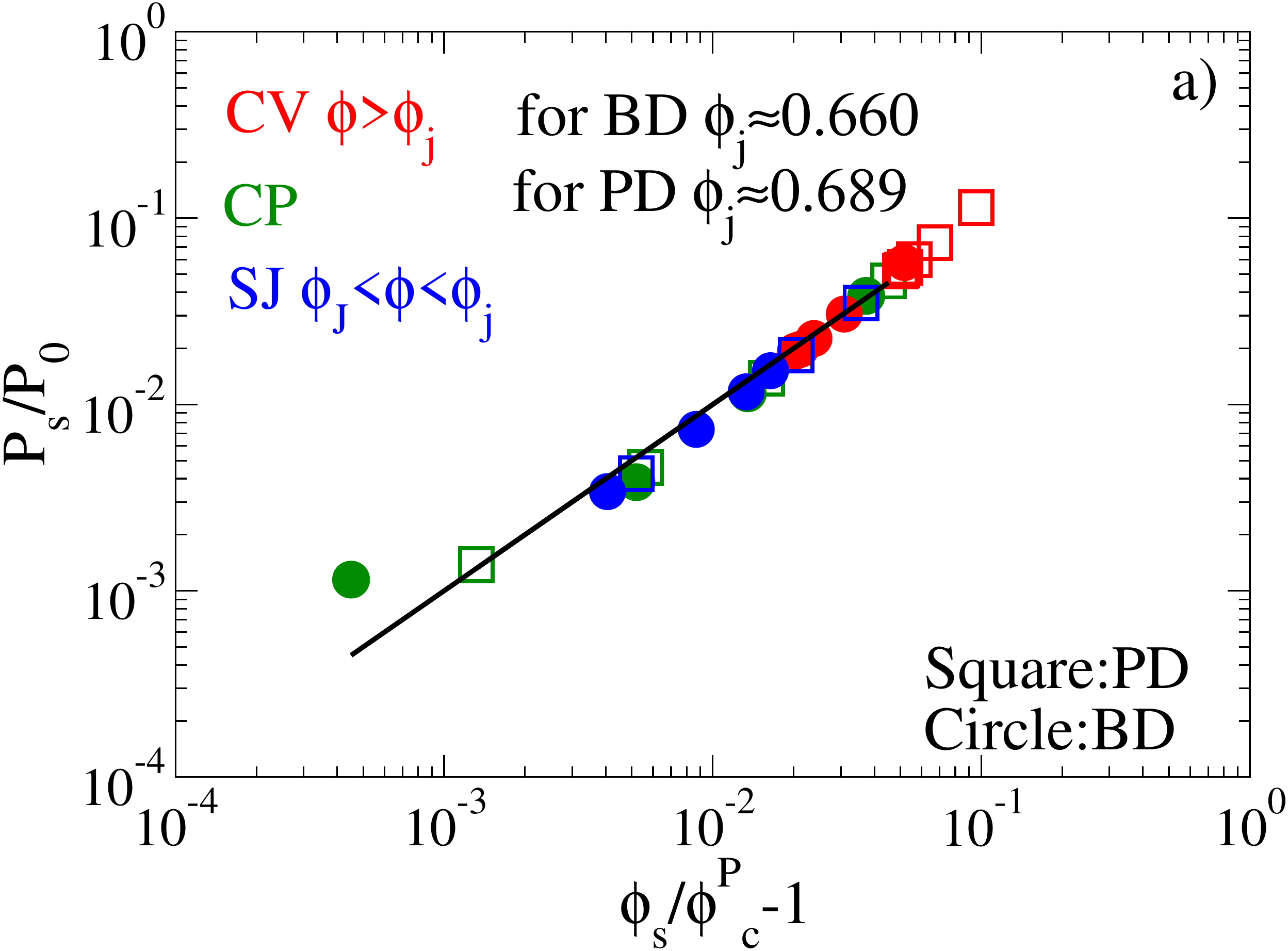}}
	\subfloat[]{\includegraphics[scale=0.34]{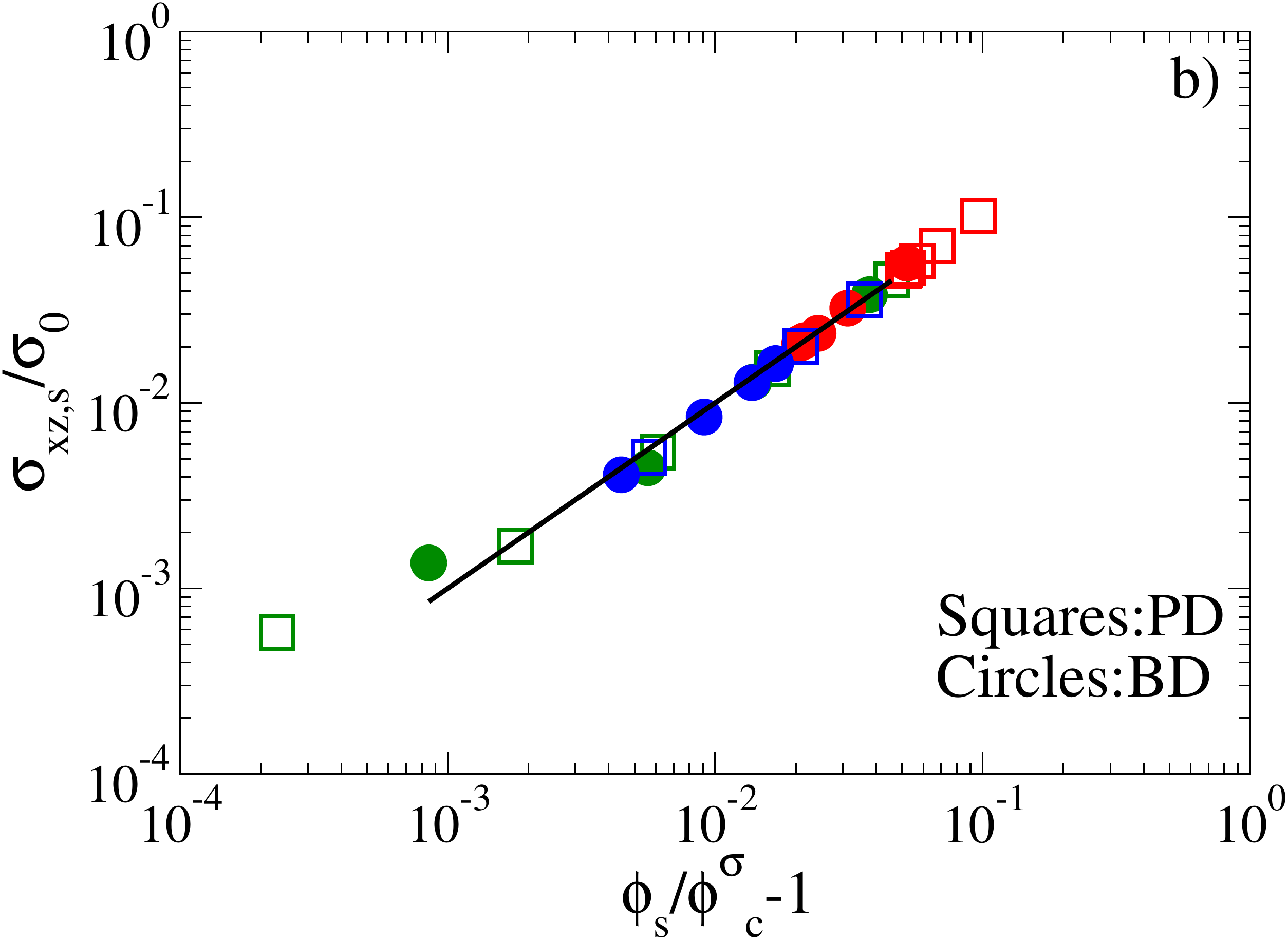}}
		\caption{\label{scaling} \textbf{
		Critical scalings of steady-states.}
		We fit the EOS data of (a) pressure and (b) stress to Eqs.~(\ref{eq:Ps}) and (\ref{eq:sigmas}). The fittings are represented by solid lines. 
		}  
\end{figure*}

 \begin{table*}
 \begin{tabular}{|c|c|c|c|c|c|c|}
     \hline
     &     $P_0$ & $\phi_c^P$ & $\sigma_0$ & $\phi_c^\sigma$ & $P_0'$ & $ \phi_J$ \\
     \hline
     BD  & 0.261 & 0.647 & 0.024 & 0.647 &0.29 & 0.648\\
    \hline 
    PD & 0.217 & 0.656 & 0.021 & 0.656 &0.21 & 0.655 \\
    \hline
 \end{tabular}
 \caption{\label{fit_values}
 Values of fitting parameters in Eqs.~(\ref{eq:Ps}), (\ref{eq:sigmas}), and (\ref{eq:Piso}), for both BD ($\phi_j = 0.660$) and PD ($\phi_j = 0.689$) models. 
 The steady-state data,  $P_0, \phi_c^P, \sigma_0$ and  $\phi_c^\sigma$, are obtained from constant pressure shear; the constant volume shear gives the same results because the EOSs are independent of shear protocols (see FIG~\ref{scaling}).
 }

\end{table*}

\subsection{Additional data for the generalized zero-temperature jamming phase diagram}
In FIG \ref{Shear_jamming} we show the generalized zero-temperature jamming phase diagram for the BD model.
Similar to the PD case (Fig.~4), the yield stress shows a discontinuous jump at $\phi_j$  for $\phi_j > \phi_J$.
This behavior is independent of the definition of the yield stress, which can be seen from Fig. 4 where $\sigma_Y$ is defined as the steady-state value $\sigma_s$, and from  FIG~\ref{yield_Stress} where $\sigma_Y'$ is defined as the peak value of the shear stress in the stress-strain curve (both figures are for the PD model).

\begin{figure*}
    \centering
    \hfill
    \subfloat{\includegraphics[scale=0.34]{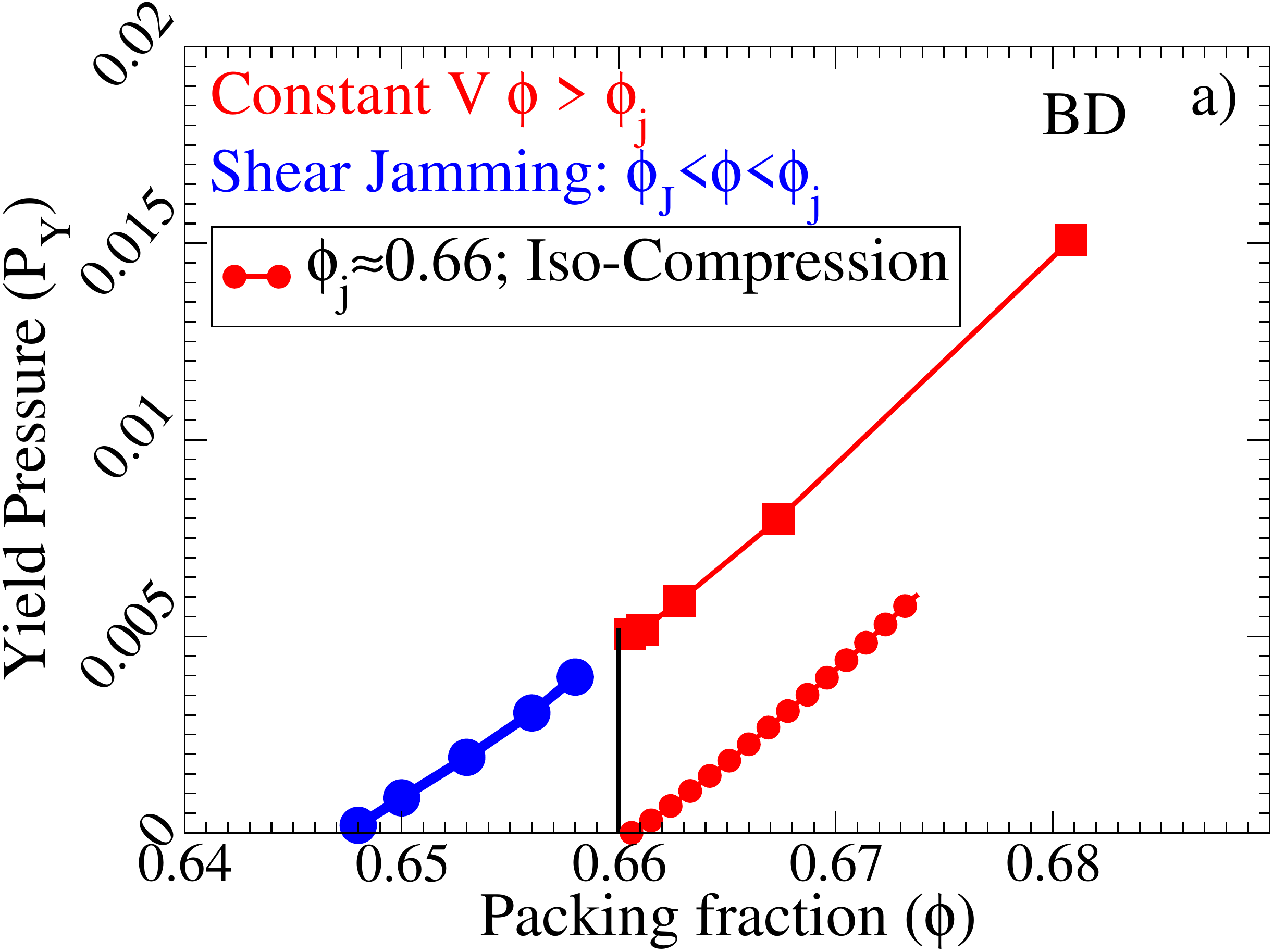}}
    \hfill
	\subfloat{\includegraphics[scale=0.34]{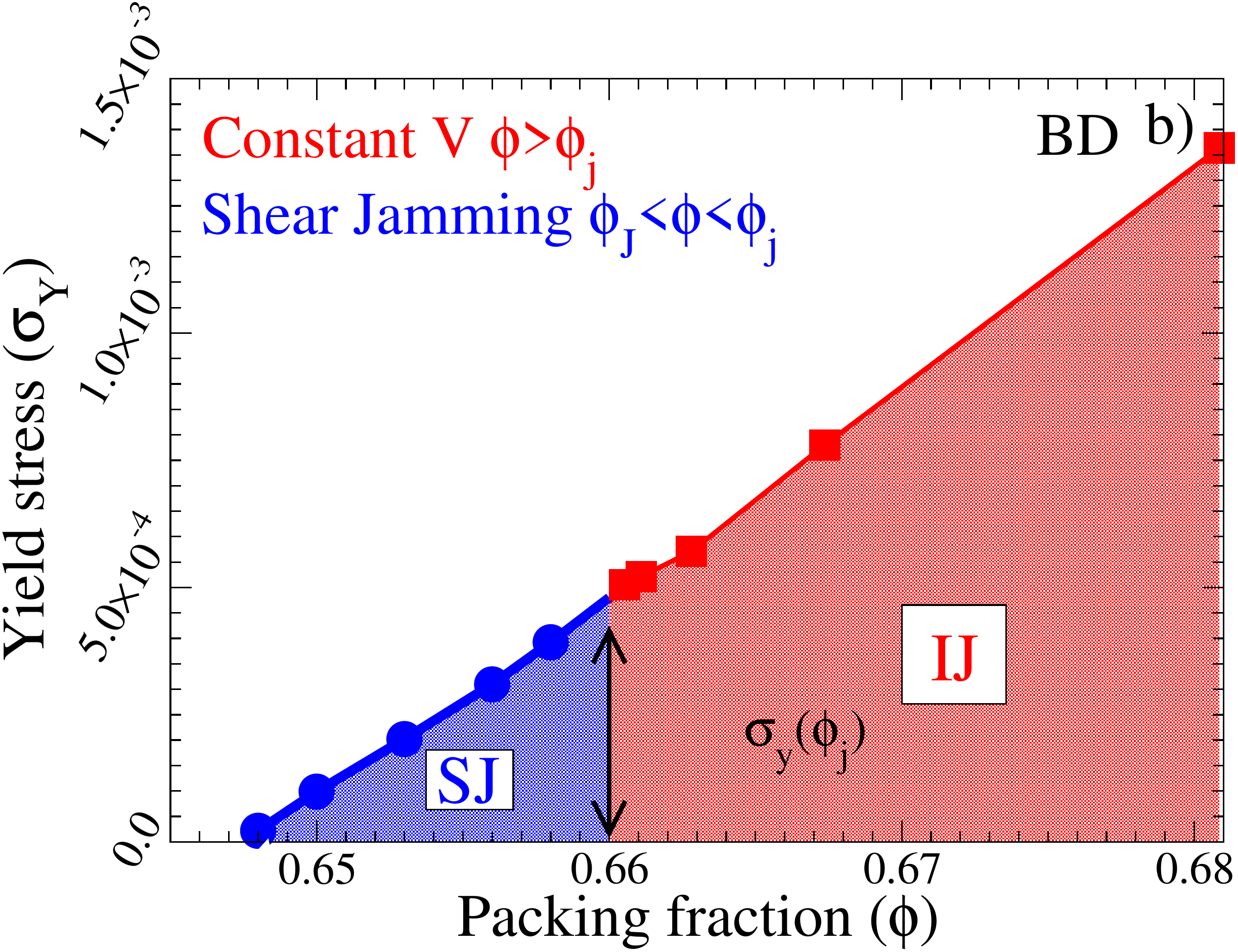}}
    \caption{\label{Shear_jamming} \textbf{Generalized zero-temperature jamming phase diagram for the BD model.}
	\textbf{a)} Yield pressure ($P_Y=P_s$) as a function of packing density $\phi$,  
	obtained by constant volume shear deformations for  both $\phi >  \phi_j=0.66$ (isotropic jamming, IJ) and $\phi <  \phi_j$ (shear jamming, SJ). 
	The isotropic compression pressure $P_{\rm iso}$ is also plotted.
	\textbf{b)} Yield stress ($\sigma_Y = \sigma_{xz,s}$) as a function of $\phi$. 
		}
    \label{fig:my_label}
\end{figure*}

\begin{figure*}
    \centering
    \includegraphics[scale=0.34]{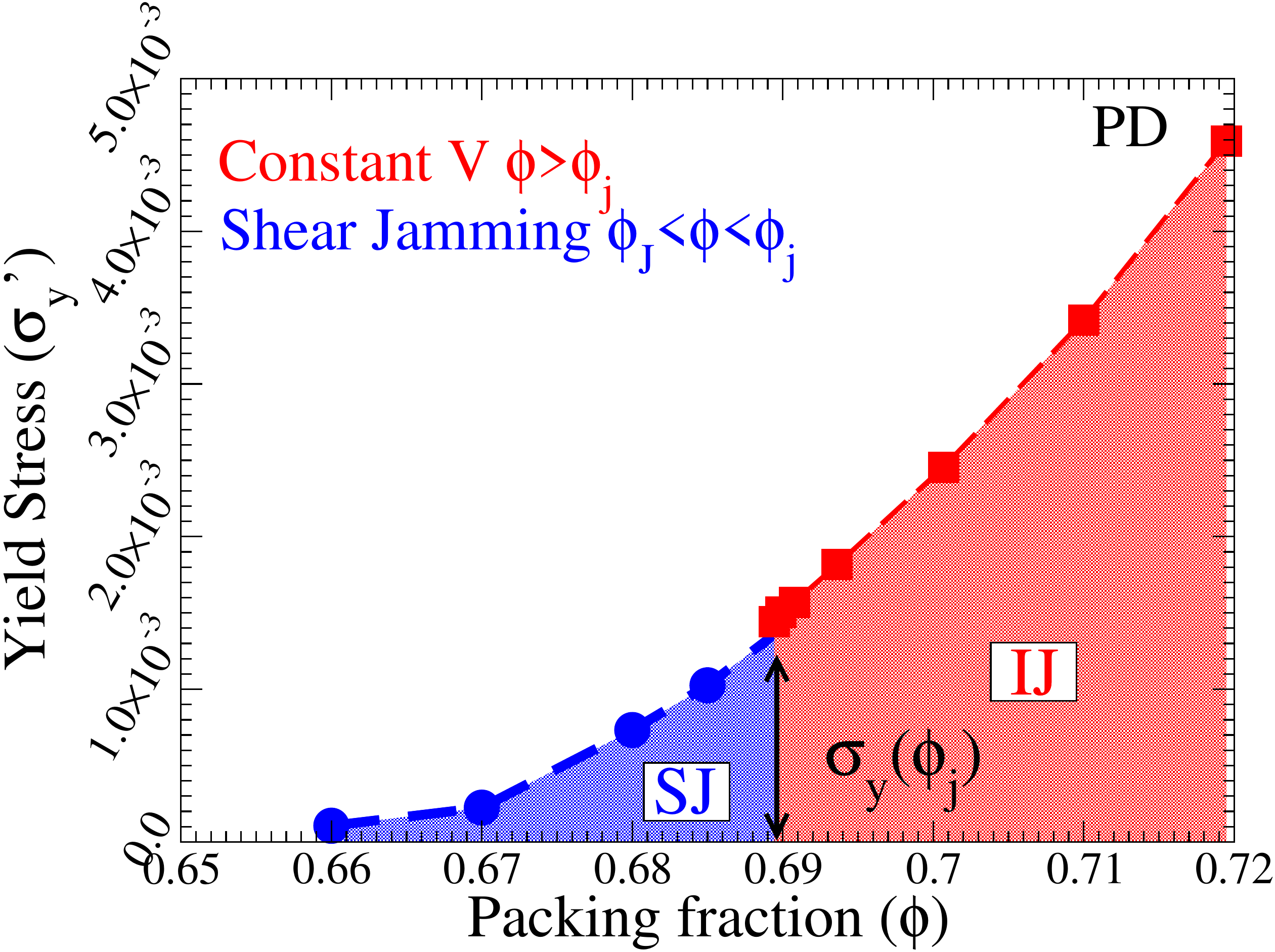}
		\caption{ \label{yield_Stress}{\bf Generalized zero-temperature jamming phase diagram for the PD model, where the yield stress $\sigma_Y'$ is defined as the peak value of the shear stress in the stress-strain curve.} The jamming density is $\phi_j = 0.69$. }
    \label{fig:my_label}
\end{figure*}




\subsection{Jamming densities of mechanically annealed bi-disperse  sphere packings} 

An over-jammed BD system at packing density $\phi$ (compressed from $\phi_J \simeq 0.647$),  unjams under constant volume cyclic AQS,
 and jams again at $\phi_j$ ($\phi_j > \phi > \phi_J$) upon a further compression. 
The jamming density $\phi_j$ 
depends on both the unjamming
density $\phi$  and the strain amplitude $\gamma_{max}$ of the cyclic shear.
As shown in the FIG~\ref{dependence_of_jamming_density}, $\phi_j$ increases with 
$\gamma_{max}$ for a fixed $\phi$, and increases with $\phi$ for a fixed $\gamma_{max}$.
 In the main text, we use $\gamma_{max}=0.07$, because for this amplitude, the largest range of densities over which unjamming occurs is obtained~\cite{pallabi}.

\begin{figure*}
	\subfloat[]{\includegraphics[scale=0.34]{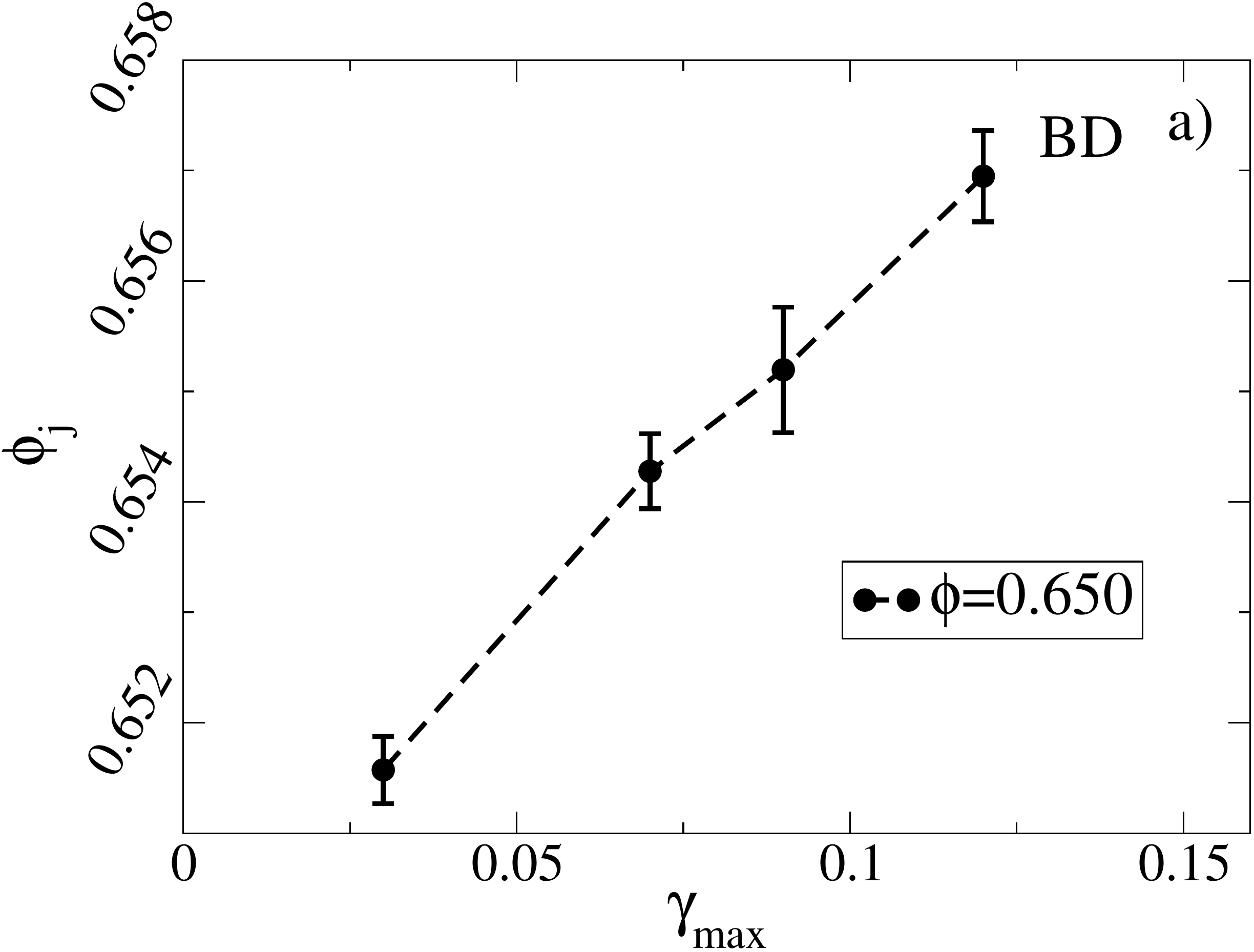}}
	\subfloat[]{\includegraphics[scale=0.34]{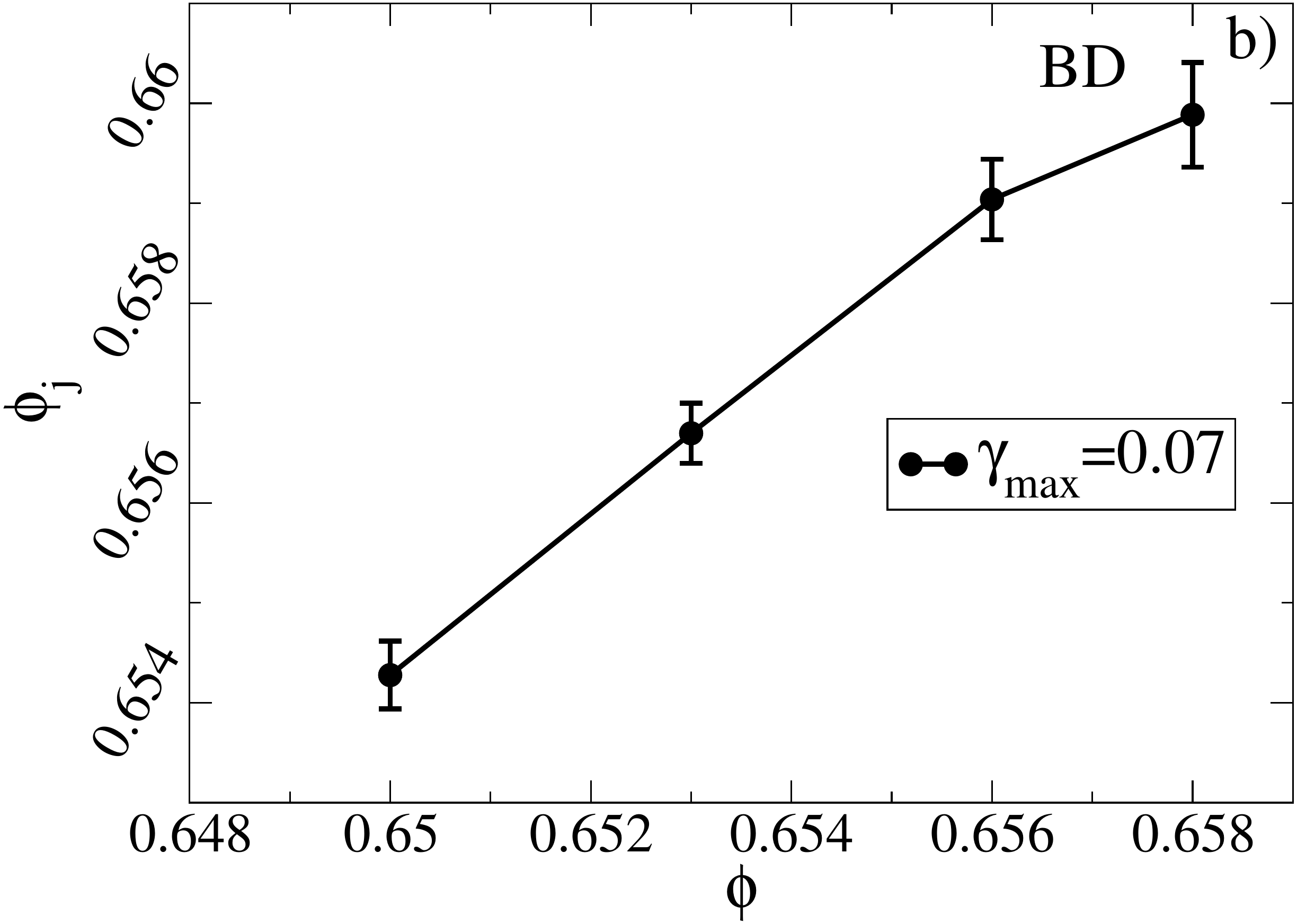}}
		\caption{ \label{dependence_of_jamming_density}\textbf{Dependence of jamming density $\phi_j$ on protocol parameters of cyclic AQS, for the BD model.} 
		\textbf{a)} Dependence of jamming density $\phi_j$ on the strain amplitude $\gamma_{max}$, for a fixed unjamming density $\phi = 0.650$. 
		\textbf{b)} Dependence of jamming density $\phi_j$ on 	the unjamming density $\phi$, for a fixed $\gamma_{max} = 0.07$.
		Error bars represent standard deviations.
		} 
\end{figure*}


\end{document}